\documentclass[sigconf]{acmart}
\AtBeginDocument{%
  \providecommand\BibTeX{{%
    \normalfont B\kern-0.5em{\scshape i\kern-0.25em b}\kern-0.8em\TeX}}}
\usepackage{graphicx,tabularx,subcaption}
\usepackage{booktabs}
\usepackage{multirow}
\usepackage[capitalise,nameinlink,compress]{cleveref}
\usepackage{colortbl}
\usepackage{xcolor}
\usepackage{comment}
\usepackage[english]{babel}
\usepackage{fontawesome5}
\usepackage{arydshln}
\usepackage{tikz}
\usepackage{rotating}
\usepackage{enumitem}
\graphicspath{{images/}}
\usepackage{framed}
\usepackage{array}

\newenvironment{frshaded*}{%
\MakeFramed {\advance\hsize-\width \FrameRestore}
}%
{\endMakeFramed}

\definecolor{main}{HTML}{fd829a}    
\definecolor{sub}{HTML}{fee2e8}     

\definecolor{lightgray}{gray}{0.9}

\definecolor{greenShade}{RGB}{240, 255, 240}
\definecolor{greenFont}{RGB}{0, 145, 107}

\definecolor{blueShade}{RGB}{240, 240, 255}
\definecolor{blueFont}{RGB}{0, 107, 179}

\definecolor{redShade}{RGB}{255, 240, 240}
\definecolor{redFont}{RGB}{255, 100, 100}

\definecolor{yellowShade}{RGB}{255, 250, 230}
\definecolor{yellowFont}{RGB}{255, 209, 81}

\definecolor{pinkShade}{RGB}{254, 236, 240}
\definecolor{pinkFont}{RGB}{251, 29, 71}

\definecolor{lightBlueShade}{RGB}{217, 243, 255}
\definecolor{lightBlueFont}{RGB}{0, 122, 176}

\definecolor{peachShade}{RGB}{255, 231, 209}
\definecolor{peachFont}{RGB}{242, 115, 0}

\definecolor{purpleShade}{RGB}{241, 232, 248}
\definecolor{purpleFont}{RGB}{112, 48, 160}

\definecolor{orangeShade}{RGB}{252, 230, 208}
\definecolor{orangeFont}{RGB}{249, 111, 7}

\copyrightyear{2025} 
\acmYear{2025} 
\setcopyright{acmlicensed}\acmConference[UIST '25]{ACM Symposium on User Interface Software and Technology}{September 28 -- October 1, 2025}{Busan, South Korea}
\acmBooktitle{Proceedings of the 38th Annual ACM Symposium on User Interface Software and Technology (UIST ’25), September 28 -- October 1, 2025, Busan, South Korea}

\begin{document}


\title[A Job Seeker's Perspective on Multi-Agent Recruitment Systems for Explaining Hiring Decisions]{Let’s Get You Hired: A Job Seeker’s Perspective on Multi-Agent Recruitment Systems for Explaining Hiring Decisions}

\author{Aditya Bhattacharya}
\orcid{0000-0003-2740-039X}
\email{aditya.bhattacharya@kuleuven.be}
\affiliation{%
  \institution{KU Leuven}
  \city{Leuven}
  \country{Belgium}
}

\author{Katrien Verbert}
\orcid{0000-0001-6699-7710}
\email{katrien.verbert@kuleuven.be}
\affiliation{%
  \institution{KU Leuven}
  \city{Leuven}
  \country{Belgium}
}

\renewcommand{\shortauthors}{Bhattacharya and Verbert}

\begin{abstract}
During job recruitment, traditional applicant selection methods often lack transparency. Candidates are rarely given sufficient justifications for recruiting decisions, whether they are made manually by human recruiters or through the use of black-box Applicant Tracking Systems (ATS). To address this problem, our work introduces a multi-agent AI system that uses Large Language Models (LLMs) to guide job seekers during the recruitment process. Using an iterative user-centric design approach, we first conducted a two-phased exploratory study with four active job seekers to inform the design and development of the system. Subsequently, we conducted an in-depth, qualitative user study with 20 active job seekers through individual one-to-one interviews to evaluate the developed prototype. The results of our evaluation demonstrate that participants perceived our multi-agent recruitment system as significantly more actionable, trustworthy, and fair compared to traditional methods. Our study further helped us uncover in-depth insights into factors contributing to these perceived user experiences. Drawing from these insights, we offer broader design implications for building user-aligned, multi-agent explainable AI systems across diverse domains.
\end{abstract}




\begin{teaserfigure}
  \centering
  \includegraphics[width=0.8\linewidth]{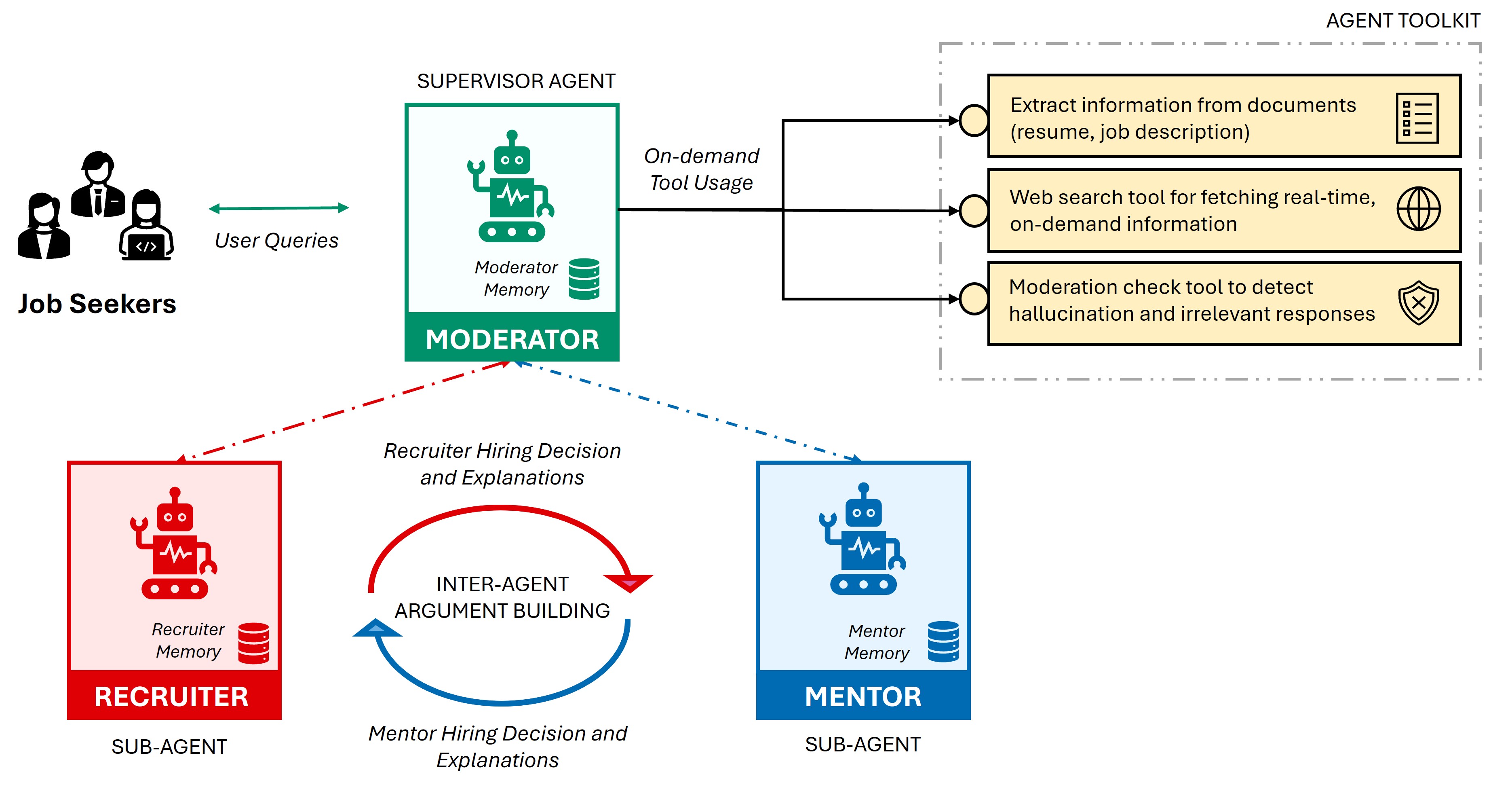}
  \caption{High-level design of our multi-agent recruitment system. The agentic architecture was designed through an iterative, user-centric process involving active job seekers, ensuring that the system aligns with their needs and expectations.}
  \Description[System Design]{High-level design of our multi-agent recruitment system. The agentic architecture was designed through an iterative, user-centric process involving active job seekers, ensuring that the system aligns with their needs and expectations.}
  \label{fig:teaser_image}
\end{teaserfigure}

\maketitle

\section{Introduction}
Despite technological advancements, the conventional job application process lacks transparency, particularly in the initial candidate shortlisting phase. Traditional approaches, whether relying on the algorithmic `black box' of Applicant Tracking Systems (ATS) or the subjective evaluations of human recruiters, often fail to provide applicants with constructive and personalised feedback for explaining the hiring decision \cite{Kochling2020}. The lack of clarity in hiring decisions negatively impacts job seekers' experiences. When decisions seem arbitrary, trust erodes, engagement declines, and uncertainty persists \cite{Kochling2020, hunkenschroer2022ethics}.

Recent advancements in Large Language Model (LLM)-based conversational AI systems have transformed the ability of decision support systems to explain their predicted outcomes \cite{caetano2025agenticworkflowsconversationalhumanai, sharma-etal-2024-investigating}. Additionally, the emergence of AI agents, i.e., autonomous systems that leverage LLMs' reasoning capabilities to perceive, analyse, and act, presents a new opportunity to increase transparency in candidate shortlisting and provide clearer explanations for hiring decisions \cite{gan2024applicationllmagentsrecruitment}. This is due to the inherent ability of agentic systems to integrate diverse data sources and perform complex reasoning of user queries, allowing for the generation of context-aware and personalised explanations \cite{Marini2023CA, fourney2024magenticonegeneralistmultiagentsolving}. For instance, in a recruitment scenario, an applicant can query an AI agent for company background information before an interview. The AI agent can utilise web search tools to retrieve relevant company details and suggest ways for the applicant to highlight their strengths during the interview, aligning with the company's core values. This capability to tailor explanations to individual user needs holds immense promise for improving the job application experience.

To effectively meet the diverse needs of end users interacting with agentic systems, prior research has emphasised the significance of multi-agent systems \cite{fourney2024magenticonegeneralistmultiagentsolving}. Unlike single AI agents, which often struggle with context switching and generate inaccurate or irrelevant responses when tasked with multiple roles, multi-agent systems distribute responsibilities among specialised agents, improving both the accuracy and relevance of the generated responses \cite{guo2024largelanguagemodelbased}. However, the effective design of these systems remains a significant challenge, particularly due to the limited user-centred research in this area  \cite{he2025plan}. Key obstacles include designing simplified system architectures with clearly defined agent roles and capabilities, establishing efficient communication protocols between agents,  and developing strategies for resolving disagreements between multiple agents  \cite{han2024llmmultiagentsystemschallenges, guo2024largelanguagemodelbased, fourney2024magenticonegeneralistmultiagentsolving, tran2025multiagentcollaborationmechanismssurvey}. 

Our research seeks to bridge these gaps by introducing a user-centric multi-agent system designed to provide actionable explanations that clarify hiring decisions for job applicants. This system enables applicants to engage in a conversational dialogue, evaluate their job profile against specific openings and obtain tailored suggestions for profile improvement and interview preparation.  We adopted an iterative, user-centred design approach (UCD) \cite{UCD2014} to develop the system. We started with a two-phased exploratory user study involving four active job seekers to capture key user requirements and co-design the system. This was followed by an in-depth qualitative study using one-to-one semi-structured interviews with 20 participants. This study focused on addressing the following research questions:

\begin{description}[topsep=0pt, itemsep=0pt]
\item[RQ1.] How do job applicants perceive the actionability of multi-agent AI feedback for improving their job applications?

\item[RQ2.] How do job applicants develop trust and confidence in a multi-agent system for hiring decisions?

\item[RQ3.] How do job applicants perceive the fairness and potential biases of multi-agent AI in hiring decisions and guidance?
\end{description}

In general, this work highlights the importance of a user-centred approach when designing multi-agent Explainable AI (XAI) systems \cite{BhattacharyaXAI2022}. Our evaluation showed that participants perceived the multi-agent recruitment system as significantly more actionable, trustworthy, and fair compared to conventional hiring methods. Additionally, the study offered valuable insights into how job seekers perceive and interpret multi-agent systems used to explain hiring decisions in comparison to conventional methods. Accordingly, this work makes three key contributions:

\begin{enumerate}[topsep=0pt, itemsep=0pt, left=0cm]
    \item  \textbf{Theoretical Contributions}: We introduce an iterative user-centric design approach specifically tailored for developing multi-agent XAI systems. Furthermore, by synthesising in-depth insights from our qualitative user study, we contribute broader design implications applicable to the creation of explainable and user-aligned multi-agent AI systems across various domains.

    \item  \textbf{Artifact Contributions}: We present a tangible instantiation of our UCD approach into a multi-agent recruitment chatbot that provides actionable feedback to job applicants. The source code, user-centric system architecture, and system prompts are open-sourced on \anon{GitHub}.
    
    \item  \textbf{Empirical Contributions}: We present findings from a qualitative evaluation with 20 active job seekers, highlighting that users perceived the system as more actionable, trustworthy, and fair compared to traditional recruitment methods. We also collected complementary quantitative data, revealing that evaluation scores for perceived actionability, trust, and fairness were significantly higher for our system than for traditional methods. These insights can inform future research on building user-centric multi-agent systems beyond the recruitment domain.
\end{enumerate}

\section{Background and Related Work}
\subsection{AI Agents}
AI agents are autonomous entities that leverage LLMs to perceive, think, and act in a certain environment to accomplish specific objectives \cite{fourney2024magenticonegeneralistmultiagentsolving, caetano2025agenticworkflowsconversationalhumanai}. These agents typically consist of three core components: (1) an LLM-based reasoning engine, (2) a suite of external tools for interfacing with external data sources (e.g., databases, APIs), and (3) a memory module for retaining contextual information \cite{han2024llmmultiagentsystemschallenges}. These systems offer distinct advantages, including automated task execution based on dynamic conditions, better decision-making through reasoned analysis, and personalised recommendations tailored to specific user contexts \cite{tran2025multiagentcollaborationmechanismssurvey, han2024llmmultiagentsystemschallenges}. 

To further improve the relevance and utility of agentic conversations, integrating techniques like Retrieval-Augmented Generation (RAG) enables the retrieval of real-time information \cite{gao2024retrievalaugmentedgenerationlargelanguage}. Moreover, a variety of prompting techniques, including ReAct \cite{yao2023react} and Chain-of-Thought (CoT) \cite{CoTPrompt2022},  are essential for assuring intended results and directing agent behaviour. These diverse strategies were selectively applied throughout the design and experimentation phases to steer agents toward producing more accurate, explainable, and user-aligned recommendations.

Prior work has shown the importance of having multiple AI agents, rather than a single agent, when designing systems to meet diverse user needs \cite{fourney2024magenticonegeneralistmultiagentsolving, han2024llmmultiagentsystemschallenges}. Single-agent systems excel at specialised tasks, while multi-agent systems coordinate multiple agents to address complex and varied tasks, where each agent fulfils a specific role.  Multi-agent architectures have demonstrated significant benefits in contexts that require collaboration, distributed knowledge, or a combination of specialised skills \cite{fourney2024magenticonegeneralistmultiagentsolving}. By distributing workloads and integrating their outputs, multi-agent systems are essential to execute tasks that exceed the capacity of individual agents.

Prior studies on AI agent system architecture reveal diverse designs, ranging from simple to complex  \cite{li2024MASarchitetcuresurvey}. These include hierarchical systems, where a central planner directs subordinate agents, decentralised networks allowing autonomous agent collaboration, and supervised structures with a lead agent overseeing sub-agents \cite{fourney2024magenticonegeneralistmultiagentsolving}. However, the design process of agentic systems by AI experts often overlooks the real needs of end-users. Consequently, this oversight leads to agentic systems that are unnecessarily complex and poorly aligned with user expectations.

Our work involves a user-centric process for designing the multi-agent architecture and profiling individual agents. This includes defining agent roles and goals, developing prompting strategies to control their behaviour, designing toolkits, managing memory, and overall user interface (UI). The system aims to deliver conversational explanations to end users by engaging them in natural, dialogue-based conversations.

\subsection{User Interfaces for Conversational Explanation}

Conversational explanations are explanations that are delivered through free-form natural language dialogues to improve user understanding of AI systems \cite{Slack2023, lakkaraju2022rethinking, zhang2024iaskfollowupquestion, luo2023providingpersonalizedexplanationsconversational, Miller2017}. Recent developments in LLM-powered AI agents have necessitated the integration of reasoning and explanation capabilities into traditional Conversational User Interfaces (CUIs), such as chatbots and voice assistants \cite{shen2023convxaideliveringheterogeneousai, Slack2023, lakkaraju2022rethinking, zhang2024iaskfollowupquestion}. Additionally, prior research has highlighted the importance of CUIs in delivering dynamic, personalised responses tailored to users’ backgrounds, needs, and preferences \cite{lakkaraju2022rethinking}.

In this work, we refer to CUIs designed for conversational explanations as Explanation-driven CUIs (xCUIs). Our research investigates how job seekers interact with an iteratively designed xCUI to understand hiring decisions and improve their job prospects. While our study focuses on recruitment, its implications extend to other domains where xCUIs can enhance the transparency of decision support systems.

\begin{figure*}
\centering
\includegraphics[width=0.85\linewidth]{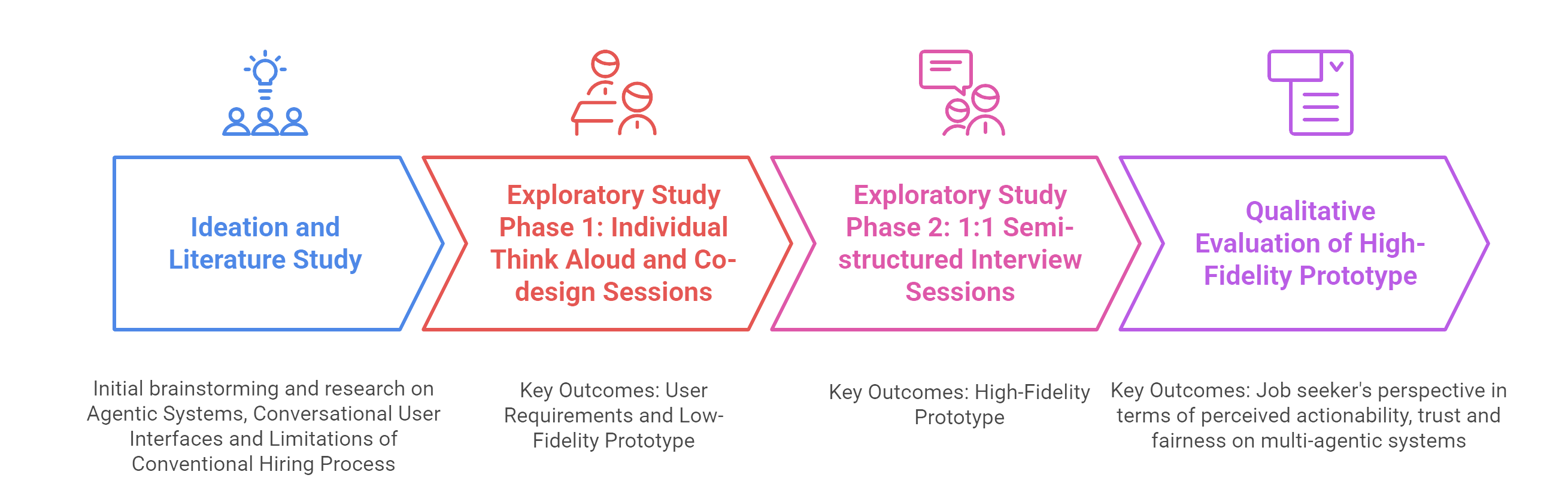}
\caption{Iterative user-centric design and development approach for our multi-agentic recruitment system.}
\Description[user-centric design approach]{Iterative user-centric design and development approach for our multi-agentic recruitment system.}
\label{fig:ucd_flow}
\end{figure*}

\section{Application Implementation: Multi Agentic Recruitment System}
\subsection{Usage Scenario}

We developed our multi-agentic xCUI using a user-centred design methodology \cite{UCD2014} (as illustrated in \Cref{fig:ucd_flow}) to support job seekers mostly in the early stages of the hiring process. The demand for xCUI systems during job recruitment is particularly high, as traditional candidate shortlisting methods are deemed to lack transparency \cite{Kochling2020}. Whether through manual assessment by human recruiters or black-box ATS platforms, candidates rarely receive adequate explanations for hiring decisions \cite{hunkenschroer2022ethics}.  This lack of transparency underscores the need for XAI solutions to increase the fairness and clarity of the decision-making process for job seekers. Furthermore, applicants often require individualised feedback to refine their resumes, acquire relevant skills, and improve their overall chances of being hired. Our xCUI system directly addresses these needs by evaluating a job seeker's profile suitability for specific openings and delivering tailored suggestions for profile enhancement and interview preparation during the hiring process.

\subsection{Exploratory User Study}
The application was implemented through an iterative UCD approach, beginning with a two-phased exploratory user study to identify key user requirements and inform the preliminary application design. This study was conducted online in two phases using Microsoft Teams (as the online meeting platform) and Miro (a real-time collaborative whiteboard). Ethical approval for the study was obtained from \anon{KU Leuven} (Approval Number: \anon{G-2024-8652-R3(MAR)}). This study involved voluntary, pro-bono participation from four active job seekers, all of whom had recent experience in the job application process. The participants were evenly distributed by gender and ranged in age from 25 to 41 years ($M=31, SD=5.1$). Recruitment was conducted via LinkedIn.

The first phase of the study involved individual think-aloud and co-design sessions. During the think-aloud sessions, we explored the key challenges job seekers face in the recruitment process and the types of support they would expect from an AI system. In the co-design component, participants engaged with existing AI tools (ChatGPT and Gemini web applications) and reimagined the design of a new support system by creating sketches on Miro. Each session lasted approximately one hour. These sessions helped us uncover participants’ core needs during the job application process and collect initial system design ideas grounded in their lived experiences.  Informed by their input and chatbot design principles from prior research \cite{yangdesigningchatbot, Weisz2024DesignApplications, he2025plan}, we created a low-fidelity prototype.  In the subsequent phase, the same four participants assessed this prototype during approximately 30-minute semi-structured interview sessions. Their feedback was crucial in refining and developing a high-fidelity prototype. Moreover, their insights extended beyond the design of the UI. We utilised their feedback to design the multi-agent system architecture and defined the specific roles and responsibilities of each agent as discussed in \cref{sec_mas_architecture}.

To facilitate qualitative analysis, all meetings have been recorded and transcribed. To derive key user requirements and inform the user-driven design of the multi-agent architecture, including agent roles and behaviour through prompt engineering, we adopted Braun and Clarke's thematic analysis method \cite{BraunClarkTA}. This approach enabled us to extract meaningful themes grounded in participants' experiences and expectations for the design and development of our system.

\subsection{User Requirements}
The following user requirements were confirmed after analysing the feedback captured from our exploratory study:

\begin{enumerate}[start=1,label={\textbf{ \arabic*.}}, topsep=0pt, itemsep=0pt, left=-0.01cm]

 \item \textbf{Provide applicants with a concise evaluation of their fit for a desired job position}: Our participants highlighted the need for a nuanced, quantitative evaluation of their resume's alignment with target job descriptions. They underscored the critical importance of an additional explanation that unpacks the rationale behind the matching score. One of the participants mentioned: ``\textit{the most frustrating aspect of job applications is the complete lack of meaningful feedback after rejections. I want to understand the real probability of my selection and receive clear, specific guidance on improving my profile. Sometimes, I feel that the opening is not an actual opening but rather for internal hiring when the rejection message is very vague}.'' This explanatory feedback would not only quantify their hiring chances but also provide actionable insights, highlighting specific strengths to leverage and precise skill gaps to address. 
 \item \textbf{Recommend necessary changes in applicant's profile to increase hiring prospects}: Participants said that, in contrast to conventional ATS-based candidate screening, they needed a system that offers practical advice on how to modify their CV or resume for a particular position in order to improve their chances of getting shortlisted. For instance, one of them mentioned: ``\textit{When transitioning between job roles, not all of my previous experience is equally important for the new role. I need the AI to tell me the most useful skills and projects that align with the new position, rather than including all of my past experiences.}''
 \item \textbf{Include a recruiter's perspective to clarify job expectations}: Participants consistently struggled with the absence of recruiter feedback, which hindered their understanding of a role's key responsibilities. They expressed the need for recruiter-like insights to evaluate their readiness, identify skill gaps, and obtain motivating feedback to better prepare for desired positions. One of them mentioned: ``\textit{Once, I was rejected thrice for a manager role, even though I knew I was a good fit. When I directly reached out to the hiring manager through LinkedIn, I was told that the HR recruiter thought I lacked prior people management experience, even though I was a senior staff engineer with people management experience. The recruiter responsible for the shortlisting process did not fully understand the role of a staff engineer. So, if the AI can think like a real recruiter, it will help to identify changes needed [in the resume] to clear the screening stage}''.
 \item \textbf{Include a mentor perspective for better interview preparation}: Participants highlighted the desired role of the AI as a mentor, suggesting it could offer career guidance by recommending suitable roles, companies, and seniority levels based on their profiles. They emphasised the need for constructive feedback delivered in a positive tone to foster confidence throughout the application process. For instance, one of them stated: ``\textit{Reaching out to a real [human] mentor is not always easy. My questions might be too basic and silly for them. Whereas the AI can mentor me and highlight my strengths so that I feel confident during interviews. I can also ask my silly questions without the fear of being judged.}''

 \item \textbf{Provide candidates with balanced comments to help them prepare for interviews}: Every participant emphasised the importance of study materials and possible interview questions, as well as assistance with interview preparation. They also voiced worries about AI systems giving answers that are excessively acceptable or positive. For instance, one of them mentioned: ``\textit{I tried ChatGPT [for evaluating their resume]. It is too positive most of the time. AI shouldn't shield us from the harsh reality, and they need to be critical sometimes to help us improve.}'' They like candid criticism that takes into account both strengths and weaknesses. Some participants proposed using a moderator to settle disputes between the viewpoints of recruiters and mentors, much like in a formal debate style: ``\textit{I like getting multiple perspectives. But to get the main [takeaway] messages, perhaps another AI can act as a debate moderator when there is a conflicting opinion between an AI mentor and an AI recruiter.}'' Such a method would guarantee a more fair and helpful assessment, giving the candidate concise and useful takeaways.

 \end{enumerate}

\subsection{Multi-Agentic System Design}\label{sec_mas_architecture}

To meet the identified user requirements, a multi-agentic system design is necessary, as participants expressed the need for the system to function simultaneously as a recruiter, mentor, and moderator. This justifies our choice of a multi-agent system over a single agent performing multiple tasks, as the latter presents several challenges, including hallucinations caused by rapid context switching \cite{tran2025multiagentcollaborationmechanismssurvey}. 
\subsubsection{Agent Roles and Responsibilities} To overcome these limitations while fulfilling user requirements, we formulated a multi-agent design consisting of the following three specialised agents, each with dedicated roles and responsibilities. The specific number of agents and the definition of these distinct roles were not based on pre-existing HR frameworks or arbitrary selection. Instead, they were carefully derived from a thorough analysis of the feedback gathered during our initial exploratory study with active job seekers. This user-centred approach ensured that the agentic composition directly aimed to address the key information needs and perspectives identified by our target users.

\begin{enumerate}[start=1,label={\textbf{ (\arabic*)}}, topsep=0pt, itemsep=0pt, left=-0.05cm]
\item \colorbox{redShade}{ \textcolor{redFont}{\textsc{\textbf{Recruiter}}}}: The recruiter agent simulates the role of a hiring manager by evaluating a candidate's suitability for a given position based on their academic background, professional experience, and relevant skill sets. It delivers objective, analytical feedback using structured scores and ratings to assess candidate-job fit. Rather than offering overly positive or generic responses, the recruiter agent focuses on constructive and actionable feedback, highlighting key areas for improvement and skill gaps. To enhance clarity and readability, its outputs are presented in a concise tabular format, avoiding long, text-heavy explanations.

\item \colorbox{blueShade}{ \textcolor{blueFont}{\textsc{\textbf{Mentor}}}}: The mentor agent functions as a supportive coach, guiding candidates in enhancing their job profiles and preparing for the recruitment process. It delivers constructive and actionable feedback by suggesting improvements for resumes, interview strategies, and skill development. In cases where recruiter feedback may appear overly critical, the mentor agent provides balancing perspectives by emphasising the candidate's strengths. Adopting a friendly, conversational, and encouraging tone, the agent provides examples and explanations to guide candidates in taking action to improve their hiring prospects.

\item \colorbox{greenShade}{ \textcolor{greenFont}{\textsc{\textbf{Moderator}}}}: The moderator agent functions as the final decision-maker, orchestrating the interaction between the recruiter and mentor agents to ensure fair, consistent, and well-reasoned hiring recommendations. It critically evaluates the perspectives offered by both agents, mediates disagreements, and provides transparent justifications for its decisions. The moderator maintains a neutral stance, avoiding overly positive or negative feedback, and summarises key points in a balanced manner to deliver an objective and comprehensive assessment to the applicant. To enhance reliability, the moderator leverages the \textit{LLM-as-a-Judge} approach \cite{zheng2023judgingllmasajudgemtbenchchatbot}, ensuring that final responses are contextually relevant and minimise hallucinations. Furthermore, the moderator agent presents its feedback in a concise and accessible format, incorporating graphical elements such as emoticons to clearly highlight strengths, gaps, and actionable next steps for the candidate.

\end{enumerate}

\subsubsection{Architectural Design Pattern} Based on insights from our exploratory study, we implemented a supervisory multi-agent architecture \cite{fourney2024magenticonegeneralistmultiagentsolving, li2024MASarchitetcuresurvey}. In this framework, a central \textsc{moderator} agent serves as the primary interface for user interactions while overseeing the \textsc{recruiter} and \textsc{mentor} agents. Although these sub-agents operate under the moderator’s supervision, they function independently, processing user queries and generating responses autonomously.
To enable constructive argumentation and ensure balanced recommendations, responses generated by the sub-agents are shared among them. This mechanism allows for counter-arguments and justifications, particularly when discrepancies arise in hiring assessments. Upon receiving a user query, the moderator determines whether to engage the recruiter agent, the mentor agent, or both, depending on the context. It then validates and synthesises the sub-agents' responses into a cohesive, structured output before presenting it to the user. \Cref{fig:teaser_image} illustrates the high-level design of our multi-agent architecture.

\subsubsection{Agent Tools} As the central supervisory agent, the moderator exclusively manages access to the system’s toolkit, a suite of functionalities designed to enhance decision-making and ensure the overall accuracy of the agentic responses. This toolkit comprises modules for extracting and interpreting information from applicant resumes and job descriptions, performing web searches to gather contextual data about the hiring company or job role, and applying moderation mechanisms to prevent hallucinations or the generation of harmful content. In contrast, the recruiter and mentor agents do not directly access these tools. This design choice was made to optimise response latency and ensure consistent, reliable tool usage. Our experiments showed that less capable LLMs, when deployed as sub-agents, were more error-prone in invoking tools. Centralising tool access within the moderator agent not only streamlined the communication pipeline but also improved the consistency and dependability of responses. Sub-agents rely on the moderator to supply the contextual information required for their individual responses. In situations where key details are unclear or incomplete, these agents can initiate clarification requests to the moderator. This coordination mechanism enables iterative refinement and facilitates the generation of more contextually grounded and refined recommendations.

\subsubsection{Agent Memory Management} While every agent maintains a log of all conversational exchanges that include the user inputs and generated responses, the moderator agent centrally manages the overall conversation history to ensure consistency and preserve contextual understanding throughout interactions. This centralised memory allows for coherent dialogues with users and seamless information flow across agents. The individual memory of sub-agents is particularly useful for generating responses to follow-up user queries that require detailed reasoning behind the recommendations. This ensures that suggestions for improving hiring prospects or interview preparation remain contextually relevant and informed by prior exchanges.

\begin{figure*}
\centering
\includegraphics[width=0.95\linewidth]{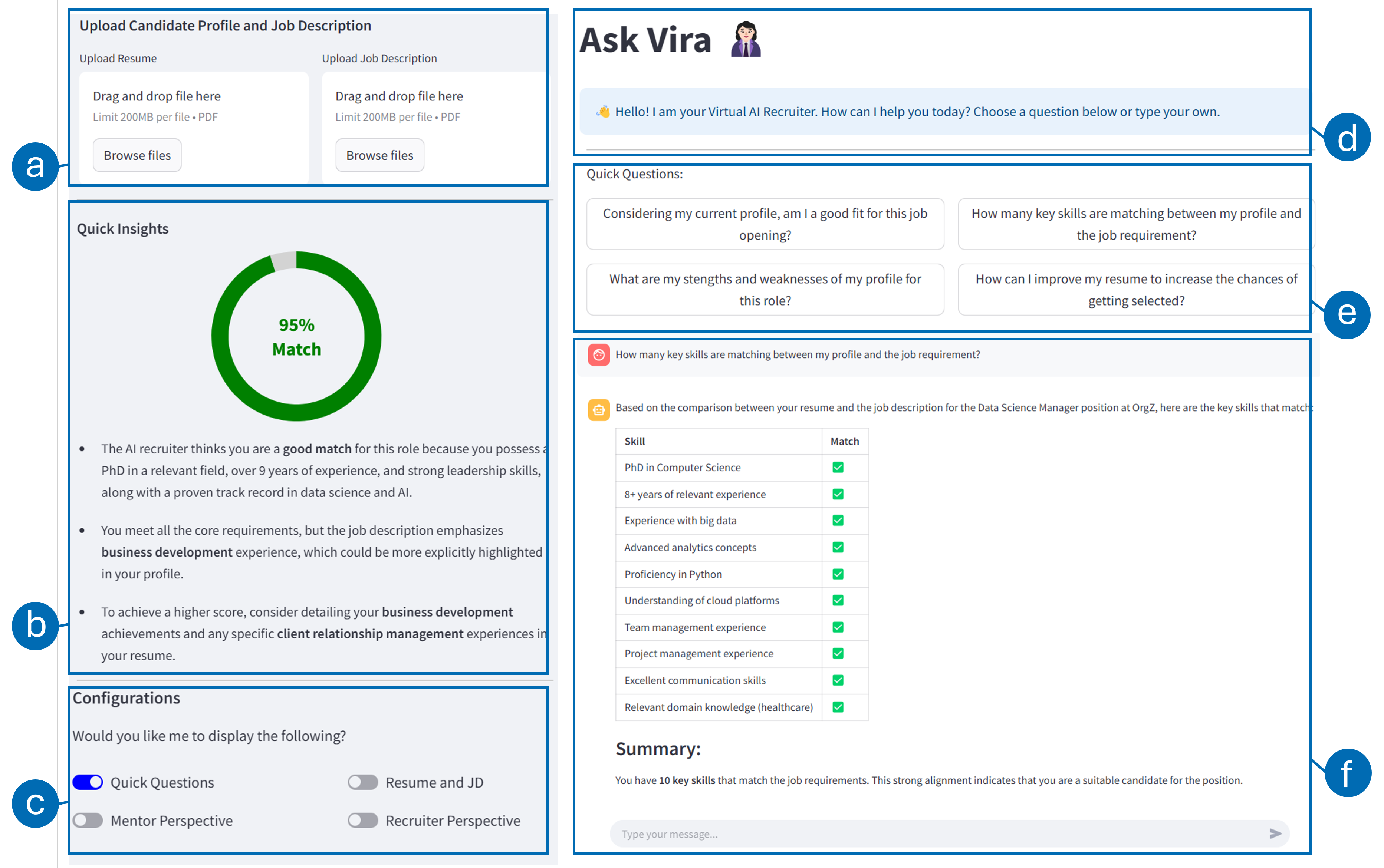}
\caption{Screenshot of our multi-agent xCUI application with the different UI components marked: (a) UI control to upload the resume and job requirement documents, (b) this component illustrates the matching score with summarised explanations, (c) configuration controls to alter the display of certain UI components, (d) personalised agent introduction, (e) quick questions to help users start the chatbot conversation and (f) agentic AI responses presented in a simplified format with necessary visual aids along with input text box for user queries.}
\Description[Multi-Agent xCUI application screenshot]{Screenshot of our multi-agent xCUI application with the different UI components marked: (a) UI control to upload the resume and job requirement documents, (b) this component illustrates the matching score with summarised explanations, (c) configuration controls to alter the display of certain UI components, (d) personalised agent introduction, (e) quick questions to help users start the chatbot conversation and (f) agentic AI responses presented in a simplified format with necessary visual aids along with input text box for user queries.}
\label{fig:app_screenshot}
\end{figure*}

\subsection{High-Fidelity Prototype}
Building on insights from our exploratory study, we developed a high-fidelity prototype of an xCUI that integrates the multi-agent system design outlined in \Cref{sec_mas_architecture}. This prototype was implemented as an interactive web application designed for job seekers navigating the recruitment process. For front-end development, we utilised Streamlit \cite{streamlit}, a Python-based framework for creating web applications. The moderator agent was powered by \textsc{GPT-4-Turbo} from OpenAI \cite{openai_online}, chosen for its advanced reasoning capabilities required to oversee and coordinate the sub-agents. Meanwhile, the recruiter and mentor agents leveraged \textsc{GPT-4o-Mini}, which offered faster response times while maintaining strong analytical performance. Additionally, we integrated LangChain \cite{langchain_online}, an open-source middleware framework that accelerates the development of LLM applications. The user interface (UI) was constructed based on an improved version of the low-fidelity prototype, incorporating feedback from the initial exploratory study. It includes matching scores accompanied by concise justifications, as well as a \textit{quick questions} feature allowing users to initiate interaction without manual typing (similar to previous research \cite{Slack2023, freed2021conversational}). The job fit score was computed as the average of the individual scores provided by each sub-agent, based on the extent to which the applicant’s experiences and qualifications aligned with the specified job requirements. Users also have the option to customise their app view to display their resume, the job description, quick questions, and a detailed overview of the individual sub-agents (recruiter and mentor) through a configuration option. \Cref{fig:app_screenshot} illustrates a screenshot of our multi-agent xCUI application.

\section{Qualitative User Study}

\subsection{Study Setup}
We conducted an in-depth qualitative user study through semi-structured interviews with 20 active job seekers. Ethical approval was obtained from the \anon{KU Leuven} Ethics Committee (approval number \anon{G-2024-8652-R3(MAR)}). The study was conducted online, with each participant participating in a one-on-one interview that lasted between 40 and 55 minutes, averaging approximately 45 minutes. Interviews were recorded and transcribed using Microsoft Teams. In addition to qualitative insights, we collected quantitative data through pre-task and post-task surveys to complement our findings. This approach allowed us to gather both qualitative and quantitative evidence to address our research questions.

\begin{table}[h]
\caption{Summarised demographic information of the qualitative user study participants.}
\label{tab:demographics_participants}
 \scalebox{.75}{
\begin{tabular}{ll}
\toprule
\textbf{Demographics} & \textbf{Participant Groups}                                                 
\\ \midrule
\multirow{3}{*}{\begin{tabular}[c]{@{}l@{}}\textsc{Age} \\ \textsc{Groups}\end{tabular}} 
    & \begin{tabular}[c]{@{}l@{}}18 - 30 years: 9 \end{tabular} \\ 
    & \begin{tabular}[c]{@{}l@{}}31 - 40 years: 6 \end{tabular} \\ 
    & \begin{tabular}[c]{@{}l@{}}41 - 50 years: 5 \end{tabular}                                                                                    
\\ \midrule
\textsc{Gender} & \begin{tabular}[c]{@{}l@{}}
Male: 11 \\
Female: 9 \\
\end{tabular}

\\ \midrule
\multirow{2}{*}{\begin{tabular}[c]{@{}l@{}}\textsc{Highest} \\ \textsc{Education} \textsc{Level}\end{tabular}} 
    & \begin{tabular}[c]{@{}l@{}}Bachelor: 13 \end{tabular} \\ 
    & \begin{tabular}[c]{@{}l@{}}Master: 7
    \end{tabular}

\\ \midrule
\multirow{2}{*}{\begin{tabular}[c]{@{}l@{}} \textsc{Overall} \\ \textsc{Work Experience}\end{tabular}} 
    & \begin{tabular}[c]{@{}l@{}} Early to Mid Career Professionals (0-5 years): 8 \end{tabular} \\ 
    & \begin{tabular}[c]{@{}l@{}}Experienced Professionals (More than 5 years): 12\end{tabular}
\\ \midrule
\multirow{2}{*}{\begin{tabular}[c]{@{}l@{}}\textsc{AI} \\ \textsc{Knowledge}\end{tabular}} 
    & \begin{tabular}[c]{@{}l@{}}No experience: 1 \\
    Novice (Knowledge of using AI tools like ChatGPT, Gemini): 15
    
    \end{tabular} \\ 
    & \begin{tabular}[c]{@{}l@{}}Intermediate (Knowledge of developing AI applications): 3 \\
    Advance (Developed AI Applications): 1
    \end{tabular}

\\ \midrule
\textsc{Country of Origin} & \begin{tabular}[c]{@{}l@{}}
United States: 9, India: 5 \\
Canada: 3, Belgium: 2, China: 1
\end{tabular}\\   

\bottomrule
\end{tabular}}
\end{table}

\subsection{Participants}
This study involved 20 participants who were active job seekers currently going through the application process. An initial cohort was recruited through LinkedIn, with subsequent participants identified via snowball sampling. To mitigate potential bias, individuals who participated in the exploratory user study were excluded from this phase. The final participant group comprised a diverse range of demographic backgrounds, including variations in gender, age, education level, country of origin, and job experience level. The summarised demographic information for all participants is presented in \cref{tab:demographics_participants}. To ensure anonymity when presenting the study's findings and quoting participants, each participant is referred to as P(N), where N represents a unique identifier from 1 to 20.

\subsection{Evaluation Measures}
To address our research questions, this section describes the various evaluation measures captured during our user study. For each evaluation measure, we captured quantitative data through validated questionnaires from prior work. Additionally, we had semi-structured interview questions to gather more in-depth qualitative feedback from the participants. We included the complete set of study questionnaires as part of the supplementary material.

\emph{Perceived Actionability}: Based on insights from our exploratory study, where participants consistently emphasised the need for actionable feedback during the candidate shortlisting and interview process, we aimed to evaluate the perceived actionability of our system. Specifically, we aimed to compare its effectiveness with conventional methods, such as ATS or manual assessments conducted by human recruiters. Building on Shoemaker et al.'s definition of actionability \cite{shoemaker2014pemat}, we define perceived actionability as the extent to which users believe the system delivers clear, feasible, and valuable actions that help them achieve their primary goals. To quantitatively measure perceived actionability, we utilised the assessment toolkit developed by Singh et al. \cite{singh2024actionabilityassessmenttoolexplainable}. Additionally, to gain qualitative insights, we designed semi-structured interview questions informed by Bhattacharya et al. \cite{Bhattacharya2023}. From the qualitative responses, we aimed to explore which aspects of the multi-agent system’s feedback users found most valuable and actionable, as well as potential improvements, particularly in the context of job recruitment.

\emph{Perceived Trust}: Given that a lack of transparency in conventional candidate shortlisting is a significant concern in the recruitment process, we aimed to evaluate how the explanations generated by our agentic system influence users' perceived trust. We adopted the definition of perceived trust in automated systems provided by Jian et al. \cite{Jian2020_trust_scale}, who define it as a user's confidence in the reliability, competence, and integrity of an automated system in providing relevant responses to achieve their goals. To quantitatively assess perceived trust, we utilised their validated questionnaire. Furthermore, our interview questions explored factors that could impact trust in multi-agent systems (informed by prior work such as \cite{MehrotraTrust2024, bhattacharya2024exmos, BhattacharyaCHI2025}) and how users might react if their trust diminished due to issues with agent responses. 

\emph{Perceived Fairness}: Due to the well-known pitfalls of LLMs \cite{fredes2024usingllmsexplainingsets, kunz-kuhlmann-2024-properties, schiller2024humanfactordetectingerrors, tjuatja-etal-2024-llms}, such as hallucinations and biased responses, we sought to assess how participants perceived the fairness of our system. Drawing from methodologies in previous studies \cite{Luo2025EARNFairness, Nakao2022}, we quantitatively evaluated the fairness of our agentic system and compared it with traditional applicant shortlisting methods. Additionally, through semi-structured interviews, we aimed to understand how users would respond when they encountered biased responses or fairness issues within the agentic system's outputs.

We also logged the conversation history for each participant to observe patterns in user queries and to evaluate the generated responses for instances of hallucination or other common issues generally observed in LLM outputs.

\begin{figure*}
\centering
\includegraphics[width=0.85\linewidth]{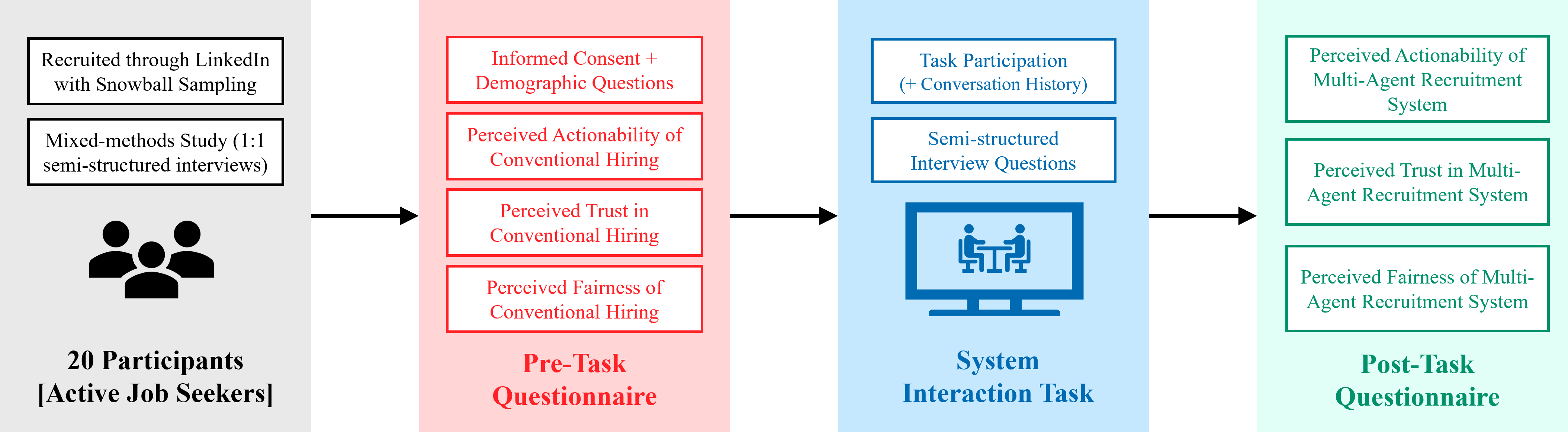}
\caption{Study flow of our qualitative user study with 20 active job seekers.}
\Description[User study flow]{Diagram illustrating the flow of our qualitative user study. The study included 20 active job seekers who were recruited through Prolific through LinkedIn with snowball sampling. The diagram also summarises evaluation measures captured through our study.}
\label{fig:user_study_flow}
\end{figure*}

\subsection{Study Procedure}
Upon recruitment, participants were provided with comprehensive information regarding the study's objectives, their roles, responsibilities, and rights. In adherence to our ethical guidelines, they were required to provide informed consent prior to participation.

Following consent, participants completed a pre-task questionnaire. This instrument served to gather demographic data and details about their prior experiences with conventional candidate shortlisting methods, whether manual (via human recruiters) or automated (via ATS). Specifically, participants were asked to reflect on their experiences in relation to our evaluation measures used to address our research questions (perceived actionability, trust, and fairness). This process allowed us to establish essential baseline scores for a direct comparison with the user experiences derived from our multi-agent xCUI system.

After completing the pre-task questionnaires, participants engaged in a task where they uploaded their resume or CV, along with a job description for a position of interest, through our application. Subsequently, the AI system evaluated their hiring chances for that specific role, displaying a matching score accompanied by justifications in the quick insights section. Participants could then interact with the chatbot to ask questions, discuss their hiring prospects and necessary resume improvements, and explore other relevant queries. Concurrent with this task, they participated in one-on-one semi-structured interviews to provide in-depth feedback on their overall experience.

After interacting with our application, each participant completed a post-task questionnaire, where they provided feedback on the system’s actionability, trustworthiness, and fairness. The quantitative data was analysed alongside qualitative interview insights to address our research questions. An overview of the study flow is illustrated in \Cref{fig:user_study_flow}.

\subsection{Data Analysis}
\emph{Quantitative Data}: For the quantitative analysis, we first evaluated the normality of our collected data using the Shapiro-Wilk test \cite{mccrum-gardner_which_2008}. Given that the data violated normality assumptions, we applied non-parametric statistical test, such as the Mann-Whitney U test \cite{mccrum-gardner_which_2008}, to compare the performance of our application against conventional methods used in job recruitment.

\noindent \emph{Qualitative Data}: We analysed the qualitative data using Braun and Clarke's six-phase thematic analysis method \cite{BraunClarkTA}. This involved an initial review of the recorded interview transcripts, followed by the generation of a list of preliminary codes derived from the data. Through several iterative cycles, we then grouped these initial codes into potential overarching themes. After carefully reviewing and refining these preliminary themes, we established a final set of themes. 
These themes provided deeper insights into the overall user experience across our evaluation measures, enabling us to address our research questions effectively. Additionally, they informed potential refinements for the next iterative development phase of our system. While the quantitative data from the 20 participants provided an exploratory overview of their perceptions, the primary focus and strength of this research lie in the rich, nuanced insights gained from the qualitative analysis of the interview data, which aimed to uncover the underlying factors driving these perceptions.

\section{Results}

\subsection{How do job applicants perceive the actionability of multi-agent AI feedback for improving their job applications? (RQ1)}
\subsubsection{Quantitative insights:} 
\begin{figure}[h]
\centering
\includegraphics[width=1.0\linewidth]{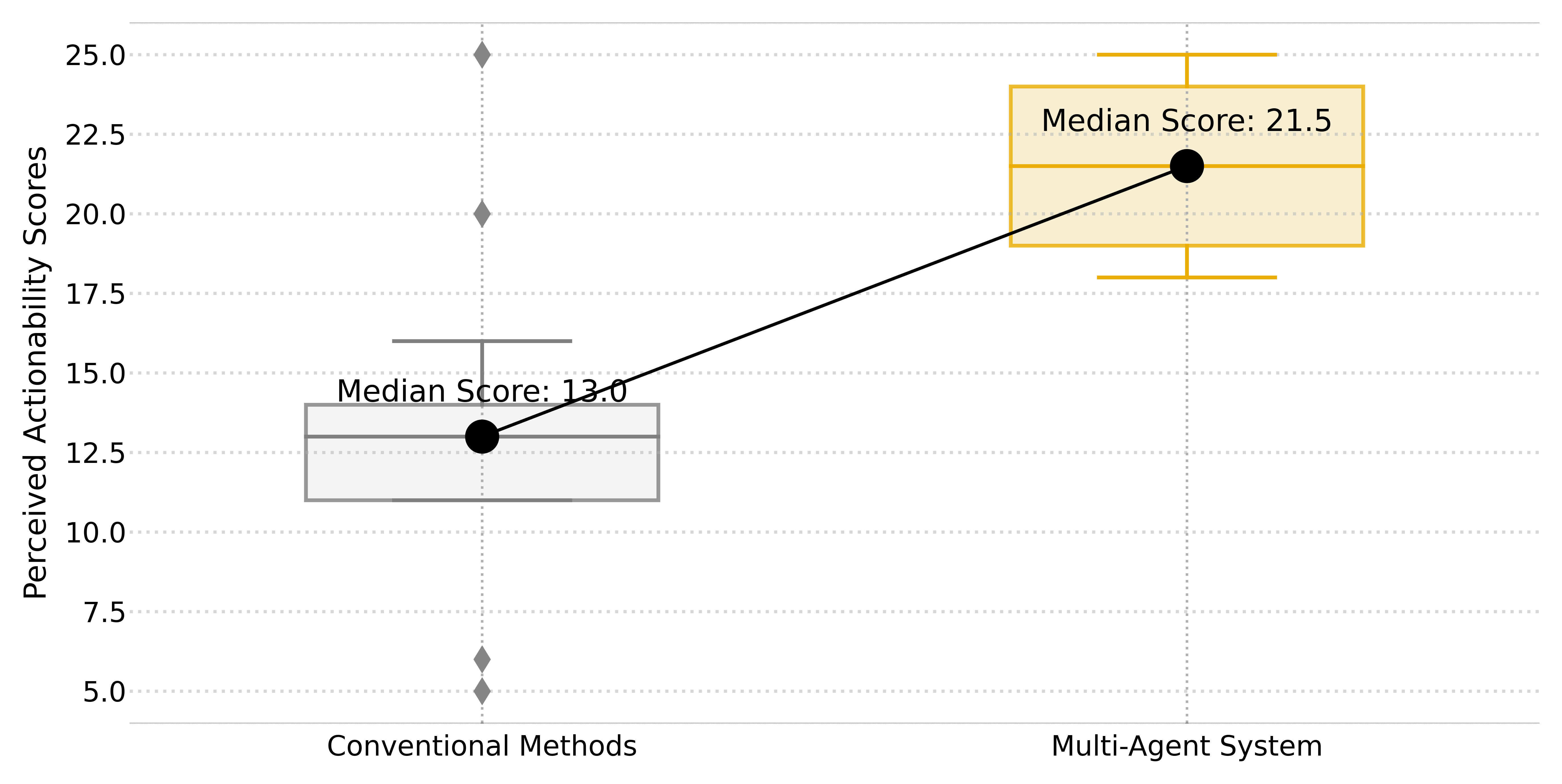}
\caption{Box plots showing an increase in the perceived actionability of our multi-agentic AI feedback compared to conventional methods}
\Description[Change in Perceived Actionability scores]{Box plots showing an increase in the perceived actionability of our multi-agentic AI feedback compared to conventional methods}
\label{fig:perceived_actionability}
\end{figure}

Our analysis revealed a statistically significant increase in perceived actionability for our system compared to participants' prior experiences with conventional methods using the Mann-Whitney U-test ($U=26, p<.001$). Specifically, as shown in \Cref{fig:perceived_actionability}, the median actionability score for our system was 34\% higher than that of conventional methods. This finding not only underscores the actionable nature of our multi-agent xCUI but also highlights a substantial lack of actionable feedback in current candidate shortlisting and recruitment processes. Our system demonstrates considerable potential to address this critical gap by providing job seekers with actionable recommendations, thereby significantly improving their overall experience during recruitment. Through qualitative data analysis, we gained a deeper understanding of job seekers' perspectives on the actionability of our system.

\subsubsection{What aspects of our multi-agent system's feedback made it seem more actionable than traditional methods?}

The following insights derived from our qualitative analysis offer a comprehensive understanding of the factors influencing the high perceived actionability of our system. Additionally, we report the number of participants supporting each identified theme.

\emph{\textbf{Improved understanding of alignment between candidate profile and job requirements}} <\textbf{16 \faUser }>: Participants found our multi-agent system's feedback more actionable due to its clear understanding of alignment between their resume and the job description of their target role. They highly valued how our system identified the most critical gaps between their current profiles and the job requirements, specifically in terms of experience and skills, and clearly outlined the necessary steps to address these discrepancies. For instance, P3 mentioned: ``\textit{It's really insightful to see how this app differentiates between must-have and good-to-have skills. Sometimes I even wonder if I still have a chance with a company if I lack a skill they're looking for. But this app clearly tells me just how important that missing skill is for the specific role and tells me how to acquire it.}''

\emph{\textbf{Personalised feedback to increase hiring prospects}} <\textbf{14 \faUser }>: Participants acknowledged the effectiveness of our system in providing personalised feedback for tailoring resumes to specific roles. This feature was particularly beneficial for candidates transitioning to new job roles, as it helped them identify transferable skills and relevant experiences while also highlighting essential skills they needed to acquire to improve their hiring prospects. For example, P2 stated: ``\textit{There are many transferable skills between data engineering and data science like knowledge of ETL and using data pipelines. This chatbot can actually help in identifying the most relevant projects I should highlight in my resume for getting selected}.'' Additionally, participants appreciated the system’s ability to recommend relevant courses and learning resources, including external web links for interview preparation:  ``\textit{Wow, this application not only tells me what I am missing in my resume, but it is suggesting online courses and videos I can watch to learn them!}'' (P1). 

\emph{\textbf{Importance of response format and xCUI components}} <\textbf{17 \faUser }>: A majority of participants appreciated the chatbot’s concise, visually enhanced summaries, which effectively highlighted their strengths, identified key gaps, and pointed out job requirements not explicitly mentioned in their resumes through the use of visual icons. As P14 noted: ``\textit{I love this concise tabular response. I can see assessment areas, recommended actions to take, their impact on the selection process, and these green ticks and red crosses that tell me which aspects to improve and which aspects I am good at.}'' Participants also emphasised the value of the \textit{quick insights} component, which enabled them to rapidly grasp the actions required and assess their overall chances of success at a glance. This feature was particularly beneficial for candidates with limited work experience (such as recent graduates), who often apply to a wide range of positions and need to quickly determine job relevance. For instance, P9 stated: ``\textit{Recent graduates like me believe in shooting a thousand arrows to hit the target once! This score and the explanations [in quick insights] are enough to give me the confidence I need to apply for openings.}'' Additionally, participants highlighted the importance of the \textit{quick questions} feature. Notably, 16 out of 20 participants initiated their interaction with the system using the provided \textit{quick questions}, while the remaining four formulated their own queries. These latter participants, who bypassed the generic suggested questions, expressed a desire for more in-depth clarification on specific gaps identified in the \textit{quick insights}.

\emph{\textbf{Importance of both mentor and recruiter perspective for increasing hiring prospects}} <\textbf{13 \faUser }>: The majority of our participants particularly valued the inclusion of both mentor and recruiter perspectives for gaining a comprehensive understanding of necessary improvements. For example, P16 mentioned: ``\textit{I like getting more perspectives as it helps me decide what is truly important and what I can de-prioritise}''. However, most of those who expressed a preference for both (9 out of 13) found the recruiter feedback more useful. They explained that the critical nature of this feedback provided a realistic assessment of their shortcomings and the level of effort required for improvement. For instance, P8 stated: ``\textit{The critical feedback is more useful. Even if it could make me feel worried, the harsh feedback can actually motivate me to take that extra step to improve my hiring chances}''. Participants who did not explicitly value the distinct perspectives of the mentor and recruiter agents frequently perceived minimal divergence in their responses, largely due to the alignment of the generated feedback. This observation suggests that the individual presentation of both mentor and recruiter perspectives may hold greater significance than exclusive reliance on a synthesised response from the moderator, particularly in instances where the sub-agents exhibit contradictory opinions. Nevertheless, the moderator fulfilled a critical function in synthesising actionable steps for the user and resolving any ambiguities arising from the sub-agents' responses.

\subsubsection{How would users react when they perceive multi-agent AI feedback as insufficiently actionable?} While the majority of participants characterised the feedback and recommendations from our multi-agent xCUI as highly actionable, a subset expressed a desire for more granular explanations (6 out of 20). This deeper curiosity centred on a more thorough understanding of identified gaps, strategies for their mitigation, and the relevance of specific skills and experiences to their target job role.  The following insights help us understand what users can do if the xCUI feedback is insufficiently actionable for them.

\emph{\textbf{Ask detailed follow-up questions for better clarification}} 
 <\textbf{6 \faUser }>: Participants who desired more detailed, granular feedback from the system initiated multiple follow-up questions. Initially, these follow-ups were often brief and sometimes vague. However, when participants perceived a need for more comprehensive explanations, their subsequent questions became well-articulated, clearly specifying the information they sought in the response. For instance, P12 stated: ``\textit{The explanations are good and useful, particularly because an ATS will never generate such explanations. However, I would want to know more about the missing 25\% score [the matching score for this participant was 75\%], not just to validate my strengths but also to prepare well to cover my weaknesses.}'' However, we also observed a reluctance among some participants (8 out of 20) to pose detailed and well-articulated questions, highlighting the potential benefit of \textit{question chaining}. This approach, where the moderator agent proactively suggests relevant follow-up questions the user might find pertinent, could address this reluctance and provide further clarification without requiring the user to formulate the queries themselves.

\emph{\textbf{Steer the agent behaviour for personalised response formats}} <\textbf{3 \faUser }>: A small proportion of participants indicated a tendency to steer the AI feedback towards greater perceived actionability. This behaviour may stem from their prior familiarity with prompt engineering techniques for LLM-based chatbots: \textit{``Since I am a person who loves more information, I will instruct the chatbot to give more details to understand why this [missing] skill is relevant and what can I do in a short time. As I've seen in ChatGPT, giving detailed instructions generates more detailed answers}'' (P17).

\subsubsection{How to make multi-agent AI feedback more actionable?} Beyond understanding why users perceived our xCUI system as actionable, we also identified several insights that could further enhance its actionability. Although these suggestions fall outside the immediate scope of this study, we include them in our results as potential future enhancements that our system could support.

\emph{\textbf{Consider personal situations for contextually robust recommendations}} <\textbf{5 \faUser }>: An interesting finding from our study highlights that contextual knowledge extends beyond a nuanced understanding of candidate profiles and job descriptions to encompass the candidate's personal situation. For instance, P5 received a system suggestion to pursue a PhD to enhance his chances of securing a senior position they were targeting. However, P5's personal and family circumstances rendered a long-term commitment to a PhD program unsuitable. This indicates the importance of incorporating personal context to refine the relevance and practicality of AI-driven recommendations. He mentioned: ``\textit{I know that getting a PhD would be nice, but considering that I have a family to take care of, a PhD at this point of my career is not an option.}'' Additionally, P3 suggested that if a particular skill is realistically unattainable, the system should explicitly indicate the negative impact of lacking that skill. This information could empower candidates to make informed decisions about whether to pursue the position or explore alternative openings that are a better fit for their current capabilities: ``\textit{If getting this skill is not possible, I would be curious to know if I'm completely out of chance for applying for this job}''.

\emph{\textbf{Provide examples of successful candidates for similar roles}} <\textbf{4 \faUser }>: Many participants expressed a desire to compare their profile and hiring prospects against individuals who were successfully recruited for comparable positions within the same company. For instance, P1 suggested: ``\textit{It would be great to know some information of candidates who successfully landed similar jobs within this company, such as the skills they had, their overall years of experience, their job title or previous employers}''. While the comprehensive data collection required to implement this feature across all companies presents a considerable challenge, individual organisations could consider maintaining a database of successful hires for specific roles. Integrating the multi-agentic system with such an internal database could effectively provide this comparative insight to job seekers.

\emph{\textbf{Simplified follow-up questioning}} <\textbf{3 \faUser }>: A few participants suggested having a UI element adjacent to each recommendation that would allow them to pose clarifying questions to the agentic system with a simple click, eliminating the need for manual typing. This feature could potentially reduce their perceived task load and enhance the overall user experience.  This feature could further improve their perceived task load and improve the overall user experience. For example, P20 stated: ``\textit{It would be cool to just select or click these suggestions and automatically a question could be asked to the chatbot for getting a more detailed answer into how to execute the recommendations}''.

\emph{\textbf{Integrate with popular external platforms}} <\textbf{3 \faUser }>: A few participants also suggested the integration of our system with external platforms commonly used by candidates to showcase their portfolios, such as LinkedIn, GitHub, and personal websites. They noted that human recruiters often review these supplementary platforms alongside resumes or CVs to gain a more comprehensive understanding of a candidate's profile. As mentioned by P2: ``\textit{It would be useful for me if this chatbot can connect to my GitHub and suggest projects from my portfolio that I should definitely include in my resume for this job}''.

\colorlet{framecolor}{yellowFont}
\colorlet{shadecolor}{yellowShade}
\setlength\FrameRule{0pt}
\begin{frshaded*}
\noindent\textbf{Key-takeaways}: Job seekers perceived the multi-agent xCUI as significantly more actionable than their prior experiences with conventional methods, attributing this to the system's ability to clearly explain profile-job alignment and provide personalised, targeted feedback. Key elements such as the structured response format with visual icons and the inclusion of both mentor and recruiter perspectives facilitated a better understanding of necessary actions for increasing their hiring prospects.
\end{frshaded*}

\subsection{How do job applicants develop trust and confidence in a multi-agent system for hiring decisions? (RQ2)}

\subsubsection{Quantitative insights:} 
\begin{figure}[h]
\centering
\includegraphics[width=1.0\linewidth]{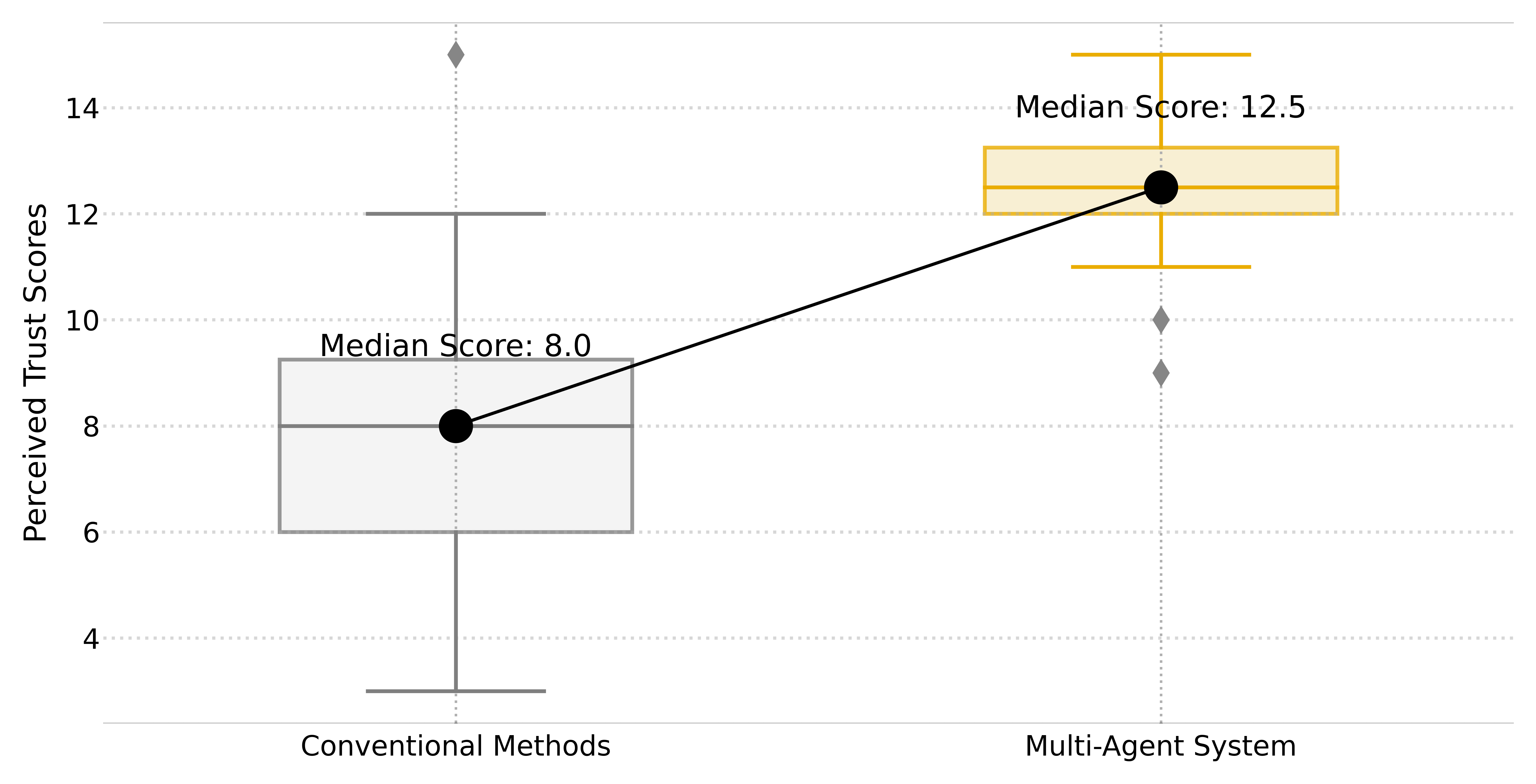}
\caption{Box plots showing an increase in the perceived trust of our multi-agentic AI feedback compared to conventional methods}
\Description[Change in Perceived Trust scores]{Box plots showing an increase in the perceived trust of our multi-agentic AI feedback compared to conventional methods}
\label{fig:perceived_trust}
\end{figure}

Our quantitative analysis of perceived trust in our multi-agent xCUI revealed significantly higher trust scores compared to participants' prior experiences with conventional hiring methods ($U=38.5, p < .001$). As shown in \Cref{fig:perceived_trust}, trust scores for our system were 30\% higher than those of conventional approaches. Furthermore, Spearman's correlation test indicated a significant positive correlation between perceived trust and perceived actionability ($r=0.54, p=.013$), suggesting that enhancing the actionability of agentic feedback can further strengthen user trust. These findings underscore the widespread lack of trust in traditional hiring processes among job seekers and demonstrate how xCUI can foster trust through transparent explanations and actionable recommendations.

\subsubsection{What are the factors that contribute to building trust?} Qualitative analysis of our user study data revealed themes that help us understand the factors contributing to building trust.

\emph{\textbf{Perceived trust rooted in understanding profile-job alignment}} <\textbf{18 \faUser }>:  The majority of participants indicated that their trust in the AI feedback stemmed from the system's ability to elucidate the alignment between their job profile and the specific requirements of the job. Participants reported that the system's assessment of their suitability and the subsequent suggestions provided often corroborated their own understanding of their qualifications in relation to the role. This congruence between the AI's evaluation and their self-assessment developed a greater sense of trust in the system's recommendations. For example, P15 stated: ``\textit{Yes, I do trust this [xCUI] feedback. The suggestions make sense; some of them are spot-on, as I am aware of these flaws. I will definitely try to follow these [recommendations].}'' This insight further explains why we observed a significant positive correlation between perceived actionability and trust.

\emph{\textbf{Multiple agentic perspective helps in building trust}} <\textbf{13 \faUser }>: Most participants expressed that feedback from both recruiter and mentor perspectives was valuable, as it increased transparency of the entire process by considering multiple aspects. For instance, P13 remarked: ``\textit{I do like both the mentor and recruiter feedback. The recruiter being more critical helps me self-reflect on what I need to improve upon. Whereas, the mentor feedback encourages me to try out the suggestions despite knowing what [skills and experiences] I lack.}'' 

\emph{\textbf{Holistic nature of agentic feedback}} <\textbf{4 \faUser }>: Several participants noted that feedback from human mentors and career coaches is often perceived as generic and influenced by personal biases or limited experiences. In contrast, they highlighted that the AI-generated feedback from our system was more specific, context-aware, and closely aligned with their individual profiles and job requirements. Participants appreciated the system’s ability to provide holistic and forward-looking recommendations, describing them as more comprehensive and pre-emptive than the typically narrow or subjective advice offered by human advisors. For example, P10 mentioned: ``\textit{The career [human] mentors whom I have ever reached out to, often gave generic feedback based on what they have seen. I understand their time is often limited when assessing my CV, but their generic feedback hasn't helped me so far. The AI feedback appears to be more specific, and I think it's more useful.}'' 

\subsubsection{How could users react when they experience less trustworthy agentic responses?} While our study did not reveal any instances of hallucinated responses (factually fabricated information), we did observe occasional short-term memory lapses within the multi-agent system, particularly during extended interactions where it appeared to forget previously captured details from candidate profiles or job descriptions. Recognising that the complete elimination of all potential issues, such as short-term memory loss, hallucinations, or irrelevant responses, may be practically challenging, we sought to understand potential user reactions to less trustworthy responses, which are summarised in the subsequent part.

\emph{\textbf{Empathy towards AI feedback}} <\textbf{18 \faUser }>: Most participants expressed empathy toward the AI system rather than disappointment or frustration, acknowledging that they did not expect the AI to be infallible or consistently accurate. For instance, P12 remarked: ``\textit{You know it's AI, I don't expect it to know everything. It's fine if it makes some mistakes, I still wouldn't mind using it.}'' Additionally, several participants noted that when the AI misinterpreted certain information, they were inclined to guide or correct it through follow-up queries, treating the interaction as a collaborative process.: ``\textit{If it [the system] fails to recognise that I’ve already mentioned a particular skill in my resume, I would explicitly point it out. Then, ask how having that skill improves my chances of getting hired}'', as stated by P15.

\emph{\textbf{Motivation towards using agentic feedback would remain same}} <\textbf{17 \faUser }>:
Most participants did not express any form of distrust toward our xCUI system. Even those who were initially more sceptical demonstrated curiosity and a willingness to engage with the system, reflecting a baseline level of confidence in its capabilities. For example, P18 mentioned: ``\textit{Even if the app tells me that I have a 100\% chance of getting selected, I would not blindly trust it. I will always trust but verify the information.}'' Interestingly, contrary to recent concerns about users’ over-reliance on LLM-generated feedback \cite{PassiVorvoreanu2022, zhai2024effects}, the majority of our participants expressed a healthy scepticism. They acknowledged the fallibility of AI and demonstrated a balanced approach toward its use. As P8 explained: “\textit{I always take AI suggestions with a pinch of salt. But it doesn't bother me if it is wrong; anyway, I won't get my hopes up even if the AI says that I am the perfect match for this role. Similarly, it'll not make me emotional if I get a poor score.}”

\colorlet{framecolor}{yellowFont}
\colorlet{shadecolor}{yellowShade}
\setlength\FrameRule{0pt}
\begin{frshaded*}
\noindent\textbf{Key-takeaways}: Job applicants demonstrated significantly greater trust in the xCUI compared to their prior experiences with conventional methods, largely attributed to the system's transparent explanations of profile-job alignment. This trust is likely sustained by the positive correlation we observed between perceived trust and actionability, suggesting that the usefulness and clarity of recommendations reinforce user trust and confidence over time. Interestingly, participants also expressed a degree of empathy towards the agentic system, acknowledging its AI nature and thus not expecting absolute error-free performance.
\end{frshaded*}

\subsection{How do job applicants perceive the fairness and potential biases of multi-agent AI in hiring decisions and guidance? (RQ3)}

\subsubsection{Quantitative insights:}
\begin{figure}[h]
\centering
\includegraphics[width=1.0\linewidth]{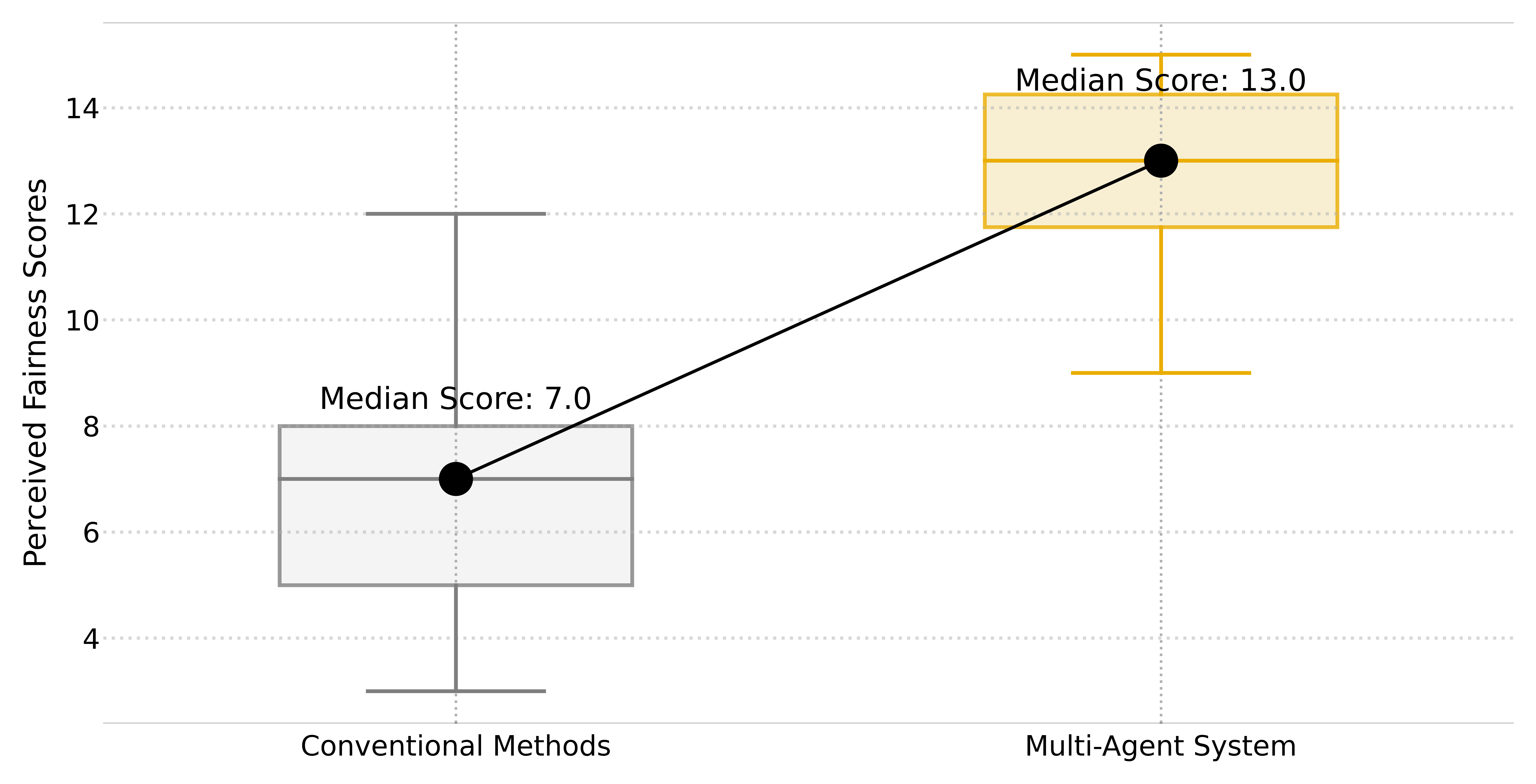}
\caption{Box plots showing an increase in the perceived fairness of our multi-agentic AI feedback compared to conventional methods}
\Description[Change in Perceived Fairness scores]{Box plots showing an increase in the perceived fairness of our multi-agentic AI feedback compared to conventional methods}
\label{fig:perceived_fairness}
\end{figure}

Our analysis of perceived fairness scores showed a significantly higher rating for our multi-agent xCUI compared to participants' prior experiences with 
 conventional hiring methods ($U=16.0, p < .001$). \Cref{fig:perceived_fairness} presents a box plot illustrating the difference in perceived fairness scores between our system and traditional recruitment approaches. We found that fairness scores for our system were 40\% higher than those of conventional methods. Moreover, Spearman's correlation test revealed a significant positive correlation between perceived fairness and trust ($r=0.52, p=.018$), as well as between perceived fairness and actionability ($r=0.67, p=.001$). These findings suggest that increasing actionability can enhance both user trust and the overall perception of fairness in a multi-agentic recruitment system.

\subsubsection{What factors contributed to the perception of our multi-agent system feedback as fairer compared to conventional methods?} The following themes help us understand why the participants deemed our system as fairer compared to conventional approaches.

\emph{\textbf{Importance of transparency and actionability of responses}} <\textbf{19 \faUser }>: Most participants described our xCUI system as transparent, noting that it provided a clear and honest picture of their hiring prospects. They appreciated that the feedback was constructive rather than overly optimistic or vague. As P19 stated: 
 ``\textit{I liked how the system is neutral. Mentioning the gaps is very helpful, and yet it does so in a neutral tone, without being rude or authoritative.}'' In addition, some participants actively tested the system for potential biases by posing sensitive or hypothetical queries. For example, P9 asked: ``\textit{Suppose I am a female applicant with two years of career break due to childbirth, would that change my selection chances?}'' The absence of biased or discriminatory responses in such cases enhanced participants’ perceptions of fairness and trustworthiness. The combination of constructive, actionable feedback and the ability to perform 'what-if' analyses contributed to participants viewing the system as both fair and transparent.

\emph{\textbf{Human bias is more concerning}} <\textbf{14 \faUser }>: A significant number of participants expressed greater concern regarding potential bias in human recruiters compared to AI systems. They suggested that human recruiters may overlook nuanced aspects of a candidate's past roles and responsibilities. Consequently, due to their narrow perspective and lack of sufficient reasoning, applicants may doubt the fairness of the recruitment process. For example, P13 mentioned: ``\textit{Sometimes it's hard for me to find out if I am getting rejected because I lack something or they [recruiters] already have a candidate from their network. So, I would say [human] recruiters are more biased and unfair than AI}''. This perspective provides insight into the observed significant difference in perceived fairness between our xCUI system and conventional recruitment approaches.

\subsubsection{How could users react if they observe fairness issues in the agentic responses?} As previously mentioned, we did not observe any instances of hallucinated responses. However, we did note evidence of short-term memory loss in the agentic system. Furthermore, we found no indications of biased responses based on demographic factors. Nevertheless, when P4 attempted to probe the system by initially uploading a job description for one role and subsequently asking questions about a related but distinct role, the agentic system continued to respond while referencing the initial role, exhibiting a hesitation to fully switch context. These observations prompted further investigation into potential user reactions upon encountering fairness issues in agentic responses.

\emph{\textbf{Steer towards the fair response through follow-up prompts without expressing any negative emotions}} <\textbf{15 \faUser }>: Consistent with user behaviour observed in response to untrustworthy outputs, participants indicated a tendency to guide the system towards fairer responses through iterative follow-up prompts when encountering potentially unfair or biased feedback. For instance, as P3 mentioned: ``\textit{I wouldn't take it [unfair response] personally knowing its AI. I don't think I will be angry. Instead, I would keep on asking questions till I get a fair answer}''. Similarly, P9 mentioned: ``\textit{Since its AI, it's meant to give you information. There is nothing to be emotional about if the information is inaccurate. I would rely on my own judgement at the end of the day.}'' These responses suggest a pragmatic user approach to perceived unfairness in agentic feedback, characterised by a willingness to refine the responses through continued interaction rather than an emotional reaction for disregarding it.

\colorlet{framecolor}{yellowFont}
\colorlet{shadecolor}{yellowShade}
\setlength\FrameRule{0pt}
\begin{frshaded*}
\noindent\textbf{Key-takeaways}: Job applicants perceived our xCUI system as significantly fairer than their prior experiences with traditional methods, attributing this to its transparent and actionable feedback. They expressed greater concern about inherent human recruiter bias as they believed that agentic systems could be designed to maintain a neutral stance. Participants also indicated a willingness to guide the system towards fairer responses even if they observed any form of bias.
\end{frshaded*}

\section{Discussion}
\subsection{Issues Beyond Hallucination: Amnestic Syndrome of LLM Agents}
As briefly discussed, we did not observe traditional hallucinations, i.e., instances where LLMs fabricate information based on unsupported assumptions \cite{xu2024hallucinationinevitableinnatelimitation}. This suggests that our hallucination-mitigation strategies, including the LLM-as-a-Judge technique, moderation tools, and limiting the context to specific information, such as candidate resumes and job descriptions, were effective. These strategies successfully filtered unreliable outputs before presenting the final moderator response. However, we acknowledge that some potential for hallucination may still exist, particularly through external web resource recommendations, which are less tightly scoped. Surprisingly, we observed a different kind of failure: short-term memory lapses, where the system temporarily forgot information explicitly mentioned earlier in the same interaction (such as specific candidate skills or job requirements). Upon reviewing conversation logs, we found that each agent correctly extracted the relevant information, but this knowledge was not consistently retained or recalled across the dialogue. We refer to this phenomenon as the \textit{amnestic syndrome of LLM agents}, a term inspired by similar patterns of temporary memory loss in human cognition. While prior work has extensively studied hallucinations in LLMs \cite{huang2023surveyhallucinationlargelanguage, xu2024hallucinationinevitableinnatelimitation}, our findings suggest that memory failures represent a distinct class of limitations in multi-agent or ExCUI-based systems. Although earlier research has begun exploring memory management strategies for LLM agents \cite{LLMMemory2024, zhang2024surveymemorymechanismlarge}, our work highlights the practical implications of amnestic syndrome of LLMs particularly for the recruitment domain. We encourage future research to investigate both the underlying causes and potential design remedies for this syndrome to ensure more reliable and coherent agentic interactions.

\subsection{Reduced Risk of Over-Reliance}

While prior research has raised concerns regarding potential user over-reliance on generative AI and LLM-based applications \cite{zhai2024effects, bo2025relyrelyevaluatinginterventions, chen2023understandingrolehumanintuition, kim2025fostering, bhattacharya2025show}, our findings suggest a comparatively lower risk of such over-reliance within the context of our multi-agentic xCUI. This reduced tendency toward blind trust could be attributed to two key factors. Firstly, a significant portion of our participants appeared to be experienced users of LLM chatbots such as ChatGPT, potentially cultivating a degree of scepticism and a heightened awareness of the inherent limitations and potential inaccuracies associated with LLM-generated content. This pre-existing understanding aligns with the concept of \textit{warranted trust}, which is posited to be crucial in fostering an appropriate calibration of trust and reliance among end-users \cite{MehrotraTrust2024}. Secondly, the design of our system, which incorporates multiple agent perspectives (i.e., recruiter, mentor, and moderator) to provide actionable explanations, likely fostered a more balanced pattern of engagement.  This multi-faceted explanation approach likely helped calibrate users’ trust and reliance as it provided diverse, sometimes contrasting, viewpoints. As also noted in prior work \cite{BrennaLi2024}, LLM applications that serve as pre-application support tools are less likely to cause over-reliance compared to systems that fully automate tasks on behalf of the user.

\subsection{Including Recruiter Perspective}
Although our xCUI system is primarily designed to support job applicants, we recognise the critical importance of incorporating recruiter perspectives in the subsequent phases of our iterative, user-centred development process. Engaging recruiters can help surface additional design opportunities and system features that cater to the needs of hiring teams, ultimately enhancing the overall utility of the platform. By involving both stakeholders (applicants and recruiters), we envision the system evolving into a more collaborative tool that facilitates mutual understanding and alignment. After all, the recruitment process should not be treated as a \textit{zero-sum game} \cite{oxford_zerosum}, where gains for one party imply losses for the other. Instead, fostering transparency and collaboration can create a \textit{positive-sum} outcome, where both applicants and recruiters benefit from more informed, efficient, and equitable hiring decisions.

\subsection{Steering of Multi-Agent Systems}

Our study highlights the proactive role that end-users are willing to take in steering the responses of multi-agent systems, particularly when the AI-generated feedback is perceived as insufficiently actionable, trustworthy, or fair. This reinforces the growing interest in user-centric control over LLM agents \cite{rahn2024controllinglargelanguagemodel, WaitGPT2024, SteerLLM2025}, with ongoing research exploring various prompting and interaction techniques to support such steering. While prior work has largely focused on single-agent systems, our findings suggest new opportunities for examining how users can effectively influence or guide multi-agentic systems. In particular, future research should explore mechanisms that allow users to selectively engage, prioritise, or reconfigure agent roles in real-time, tailoring system behaviour to meet individual needs and contextual demands more effectively.

\subsection{Ethical Considerations for Agentic Systems}

Even though participants in our study did not raise explicit concerns regarding fairness, bias, or trust in our agentic system, our user study primarily explored potential user reactions to problematic responses in a research setting. In real-world deployments, users may express reservations about being evaluated by AI agents, especially in high-stakes contexts that can impact life or livelihood \cite{diamond2024ethicalconsiderationsgenerativeagents}. Additionally, concerns around data privacy and security of user information remain paramount \cite{diamond2024ethicalconsiderationsgenerativeagents}. To promote the responsible use of agentic systems, we advocate for the implementation of informed consent mechanisms, ensuring users are aware of how their data is used, stored and safeguarded. Such systems must also comply with relevant regulatory frameworks, such as the GDPR, the EU AI Act, and similar policies, which emphasise accountability, transparency, and risk mitigation. Furthermore, while we recognise the research value of enabling user-driven steering in agentic systems, this functionality also opens potential vectors for misuse, including adversarial attacks like prompt injection and jailbreak attempts \cite{liu2024promptinjectionattackllmintegrated}. As the use of multi-agent systems becomes more prevalent, we encourage future work to not only advance their capabilities but also develop robust safeguards and ethical frameworks to ensure their safe and responsible deployment.

\subsection{Limitations}
Before discussing the broader implications of our work across multiple domains, we would like to acknowledge the following known limitations:
\begin{enumerate}[start=1,label={ (\arabic*)}, left=0.1cm]
    \item \textit{Limited evaluation across diverse LLMs for agentic system design}: Our implementation of the multi-agentic system exclusively relied on OpenAI's GPT models as the underlying LLM-based reasoning engines. While these models have demonstrated strong performance in various tasks, we acknowledge that different LLMs may exhibit varying behaviours, strengths, and limitations depending on the task and domain. Recent research has increasingly emphasised the importance of carefully selecting LLMs for specific use cases, as model choice can significantly impact response quality, alignment, and reasoning capabilities \cite{aydin2025generativeaiacademicwriting, safavinaini2024visionlanguagelargelanguagemodel,ahmad2024largescalemoralmachineexperiment}. Future work should include a comparative empirical analysis of multiple LLMs to evaluate their suitability for different agent roles within a multi-agentic architecture.
    
    \item \textit{Latency in multi-agent response generation}:
    For certain queries, we observed that the response time of our multi-agent system extended up to a minute. Such latency may negatively affect the user experience, particularly in real-time conversational settings where users expect prompt responses. This delay was often a result of some agents \textit{over-thinking} on certain user queries requiring in-depth reasoning. This syndrome can be considered similar to the  \textit{analysis paralysis syndrome} in humans \cite{analysisparalysis}, in which over-thinking restricts from taking prompt actions. Similar properties of LLM agents have also been noted in prior work \cite{christakopoulou2024agentsthinkingfastslow}. Future research should investigate the user experience implications of these delayed responses from over-thinking LLM agents, using tools such as NASA-TLX \cite{HART1988139} or through comprehensive qualitative evaluations.

    \item \textit{Potential presence of subjective recall bias in our quantitative comparisons}: In our study, participants were not exposed to a standardised ``conventional hiring system.'' Instead, their comparisons were based on personal memory and prior experiences with traditional hiring practices. As a result, our quantitative findings may be influenced by subjective recall bias \cite{stone2002capturing}. While our qualitative findings provide rich context that supports and explains these perception-based trends, the comparative feedback should not be interpreted as objective performance data. Future work could involve controlled A/B testing or randomised between-subject studies to more rigorously assess the perceived and actual effectiveness of multi-agent recruitment systems at scale.

\end{enumerate}

\subsection{Design Implications Considering Broader Applicability}
Although our work focuses specifically on the recruitment domain, the following design implications derived from our findings offer broader applicability across diverse domains involving multi-agent systems.

\begin{itemize}[left=0.1cm]
    \item \textit{User-centric design of multi-agent architectures}: Our findings emphasise the value of adopting a user-centred design approach in developing complex multi-agent system architectures. We advocate for involving end-users early in the design process to collaboratively determine the number of agents, their distinct roles and behaviours, appropriate communication protocols, and memory management strategies. An iterative user-centred development process ensures that the system aligns closely with real-world user needs.
    \item \textit{UI elements to support actionable conversations}: Findings from our study advocate the importance of simplified user interactions for having engaging and actionable conversations. We found the importance of the \textit{quick questions} and \textit{quick insights} components, especially for initiating conversations. We further recommend implementing \textit{question chaining}, either via agent-generated follow-ups or interactive UI components to scaffold deeper conversations (e.g., exploring \textit{how-to}, \textit{why}, or \textit{what-if} queries) without requiring users to manually articulate complex follow-up questions, which can often be cognitively taxing.    \item \textit{Adaptive abstraction of agentic responses}: Our study highlighted a divergence in user preferences regarding the level of detail in agentic responses. While some participants appreciated granular explanations, others found overly detailed outputs overwhelming and difficult to process. This variation underscores the challenge of determining a one-size-fits-all approach to explanation depth. To accommodate diverse cognitive preferences, we recommend integrating UI controls that allow users to dynamically adjust the level of detail through simple visual aids, such as icons or buttons. This feature can allow users to easily modify the level of detail in subsequent interactions with a single action, fostering a more personalised and cognitively accessible interaction experience.
    \item \textit{Inclusion and validation of situational information of end-users for producing context-aware recommendations}: As observed in our study findings, awareness of end-users' situational information is crucial for generating context-aware recommendations. For example, participants suggested incorporating information regarding their willingness to pursue higher education, the practical feasibility of undertaking side projects, certification courses, or even internships. Considering these personal constraints can lead to more realistic and useful recommendations. Therefore, we recommend proactively gathering such situational information through a pre-questionnaire, which could be implemented via interactive UI elements or through direct questioning by the chatbot prior to recommendation generation. Furthermore, drawing an analogy to moderation check tools, we suggest integrating agent tools to validate the alignment of final agentic recommendations with the end-users' stated situational information.
    \item \textit{Integration with external platforms via RAG for enhanced contextual knowledge}: Some participants from our user study expressed a need for the system to integrate with external platforms for information retrieval. We posit that in other application domains, such as healthcare, finance, and education, accessing specific, validated information from pre-defined sources could be even more critical for facilitating robust context-aware conversations. Therefore, we recommend implementing Retrieval-Augmented Generation (RAG) pipelines \cite{gao2024retrievalaugmentedgenerationlargelanguage} to establish a relevant and reliable contextual knowledge base derived from pertinent external platforms such as LinkedIn, GitHub, and other personal websites. Integrating this knowledge base as a tool for the LLM agents would enable more accurate information retrieval and the synthesis of highly context-aware responses.
    \item \textit{UI indicators for warranted trust and response latency in over-thinking agents}: 
    While our study found a low risk of over-reliance on the multi-agentic xCUI system in the recruitment domain, this may not generalise to other high-stakes contexts such as healthcare or finance, where users might place unwarranted trust in AI-generated outputs. Drawing from prior research on calibrated trust and transparency \cite{MehrotraTrust2024, kim2025fostering}, we recommend incorporating visual cues within the interface, such as warning messages, icons, or tooltips, to proactively alert users to potential limitations in the agents’ reasoning or coverage. Additionally, in cases where agents require extended processing time due to complex reasoning tasks (i.e., \textit{over-thinking}), explicit latency indicators or progress animations can help manage user expectations and reduce frustration. These interface features would support more informed and trust-calibrated interactions across diverse usage scenarios.
    
\end{itemize}

\section{Conclusion}
In this work, we introduced a multi-agent conversational AI system designed to support job applicants with transparent and actionable feedback during the recruitment process. Through an iterative user-centred design approach and a qualitative study involving 20 active job seekers, we demonstrated that our system was perceived as more trustworthy, fair, and actionable compared to conventional hiring methods. Our findings highlight the potential of multi-agent LLM systems in enhancing user experience, not only by offering diverse perspectives but also by providing actionable, trustworthy and fair feedback. Beyond the recruitment domain, we provide broader design implications for building explainable, user-aligned multi-agent systems across multiple application domains.

\begin{acks}
Many thanks to Yizhe Zhang, Grzegorz Meller, Robin De Croon and Maxwell Szymanski for their helpful suggestions that improved this work. We also extend our gratitude to all the participants for taking part in our user studies. This research was supported by the Researcher Access Program from Open AI, Flanders AI Research Program (FAIR) and Research Foundation–Flanders (FWO grants G0A4923N and G067721N)~\cite{BhattacharyaCHIDC, Bhattacharya2024HowGoodIsYourExplanation}.
\end{acks}

\bibliographystyle{ACM-Reference-Format}
\bibliography{references}


\begin{thebibliography}{71}


\ifx \showCODEN    \undefined \def \showCODEN     #1{\unskip}     \fi
\ifx \showDOI      \undefined \def \showDOI       #1{#1}\fi
\ifx \showISBNx    \undefined \def \showISBNx     #1{\unskip}     \fi
\ifx \showISBNxiii \undefined \def \showISBNxiii  #1{\unskip}     \fi
\ifx \showISSN     \undefined \def \showISSN      #1{\unskip}     \fi
\ifx \showLCCN     \undefined \def \showLCCN      #1{\unskip}     \fi
\ifx \shownote     \undefined \def \shownote      #1{#1}          \fi
\ifx \showarticletitle \undefined \def \showarticletitle #1{#1}   \fi
\ifx \showURL      \undefined \def \showURL       {\relax}        \fi
\providecommand\bibfield[2]{#2}
\providecommand\bibinfo[2]{#2}
\providecommand\natexlab[1]{#1}
\providecommand\showeprint[2][]{arXiv:#2}

\bibitem[Aydin et~al\mbox{.}(2025)]%
        {aydin2025generativeaiacademicwriting}
\bibfield{author}{\bibinfo{person}{Omer Aydin}, \bibinfo{person}{Enis Karaarslan}, \bibinfo{person}{Fatih~Safa Erenay}, {and} \bibinfo{person}{Nebojsa Bacanin}.} \bibinfo{year}{2025}\natexlab{}.
\newblock \bibinfo{title}{Generative AI in Academic Writing: A Comparison of DeepSeek, Qwen, ChatGPT, Gemini, Llama, Mistral, and Gemma}.
\newblock
\newblock
\showeprint[arxiv]{2503.04765}~[cs.CY]
\urldef\tempurl%
\url{https://arxiv.org/abs/2503.04765}
\showURL{%
\tempurl}


\bibitem[Bhattacharya(2022)]%
        {BhattacharyaXAI2022}
\bibfield{author}{\bibinfo{person}{Aditya Bhattacharya}.} \bibinfo{year}{2022}\natexlab{}.
\newblock \showarticletitle{Applied Machine Learning Explainability Techniques}.
\newblock In \bibinfo{booktitle}{\emph{Applied Machine Learning Explainability Techniques}}. \bibinfo{publisher}{Packt Publishing}, \bibinfo{address}{{Birmingham, UK}}.
\newblock
\showISBNx{978-1803246154}
\urldef\tempurl%
\url{https://www.packtpub.com/product/applied-machine-learning-explainability-techniques/9781803246154}
\showURL{%
\tempurl}


\bibitem[Bhattacharya(2024)]%
        {BhattacharyaCHIDC}
\bibfield{author}{\bibinfo{person}{Aditya Bhattacharya}.} \bibinfo{year}{2024}\natexlab{}.
\newblock \showarticletitle{{Towards Directive Explanations: Crafting Explainable AI Systems for Actionable Human-AI Interactions}}. In \bibinfo{booktitle}{\emph{Extended Abstracts of the CHI Conference on Human Factors in Computing Systems (CHI EA '24)}} (Honolulu, HI, USA) \emph{(\bibinfo{series}{CHI EA '24})}. \bibinfo{publisher}{ACM}, \bibinfo{address}{New York, NY, USA}, \bibinfo{pages}{6}.
\newblock
\urldef\tempurl%
\url{https://doi.org/10.1145/3613905.3638177}
\showDOI{\tempurl}


\bibitem[Bhattacharya et~al\mbox{.}(2023)]%
        {Bhattacharya2023}
\bibfield{author}{\bibinfo{person}{Aditya Bhattacharya}, \bibinfo{person}{Jeroen Ooge}, \bibinfo{person}{Gregor Stiglic}, {and} \bibinfo{person}{Katrien Verbert}.} \bibinfo{year}{2023}\natexlab{}.
\newblock \showarticletitle{Directive Explanations for Monitoring the Risk of Diabetes Onset: Introducing Directive Data-Centric Explanations and Combinations to Support What-If Explorations}. In \bibinfo{booktitle}{\emph{Proceedings of the 28th International Conference on Intelligent User Interfaces}} (Sydney, NSW, Australia) \emph{(\bibinfo{series}{IUI '23})}. \bibinfo{publisher}{Association for Computing Machinery}, \bibinfo{address}{New York, NY, USA}, \bibinfo{pages}{204–219}.
\newblock
\showISBNx{9798400701061}
\urldef\tempurl%
\url{https://doi.org/10.1145/3581641.3584075}
\showDOI{\tempurl}


\bibitem[Bhattacharya et~al\mbox{.}(2025a)]%
        {BhattacharyaCHI2025}
\bibfield{author}{\bibinfo{person}{Aditya Bhattacharya}, \bibinfo{person}{Simone Stumpf}, \bibinfo{person}{Robin~De Croon}, {and} \bibinfo{person}{Katrien Verbert}.} \bibinfo{year}{2025}\natexlab{a}.
\newblock \showarticletitle{Explanatory Debiasing: Involving Domain Experts in the Data Generation Process to Mitigate Representation Bias in AI Systems}. In \bibinfo{booktitle}{\emph{CHI Conference on Human Factors in Computing Systems (CHI '25)}} (Yokohama, Japan). \bibinfo{publisher}{ACM}, \bibinfo{address}{New York, NY, USA}, \bibinfo{pages}{20}.
\newblock
\urldef\tempurl%
\url{https://doi.org/10.1145/3706598.3713497}
\showDOI{\tempurl}


\bibitem[Bhattacharya et~al\mbox{.}(2024)]%
        {bhattacharya2024exmos}
\bibfield{author}{\bibinfo{person}{Aditya Bhattacharya}, \bibinfo{person}{Simone Stumpf}, \bibinfo{person}{Lucija Gosak}, \bibinfo{person}{Gregor Stiglic}, {and} \bibinfo{person}{Katrien Verbert}.} \bibinfo{year}{2024}\natexlab{}.
\newblock \showarticletitle{{EXMOS: Explanatory Model Steering Through Multifaceted Explanations and Data Configurations}}. In \bibinfo{booktitle}{\emph{Proceedings of the CHI Conference on Human Factors in Computing Systems}} (Honolulu, HI, USA) \emph{(\bibinfo{series}{CHI '24})}. \bibinfo{publisher}{Association for Computing Machinery}, \bibinfo{address}{New York, NY, USA}.
\newblock
\showISBNx{979840070330}
\urldef\tempurl%
\url{https://doi.org/10.1145/3613904.3642106}
\showDOI{\tempurl}


\bibitem[Bhattacharya et~al\mbox{.}(2025b)]%
        {bhattacharya2025show}
\bibfield{author}{\bibinfo{person}{Aditya Bhattacharya}, \bibinfo{person}{Tim Vanherwegen}, {and} \bibinfo{person}{Katrien Verbert}.} \bibinfo{year}{2025}\natexlab{b}.
\newblock \showarticletitle{``Show Me How'': Benefits and Challenges of Agent-Augmented Counterfactual Explanations for Non-Expert Users}. In \bibinfo{booktitle}{\emph{Proceedings of the 33rd ACM Conference on User Modeling, Adaptation and Personalization (UMAP '25)}} (New York City, NY, USA). \bibinfo{publisher}{ACM}, \bibinfo{pages}{11 pages}.
\newblock
\urldef\tempurl%
\url{https://doi.org/10.1145/3699682.3728321}
\showDOI{\tempurl}


\bibitem[Bhattacharya and Verbert(2024)]%
        {Bhattacharya2024HowGoodIsYourExplanation}
\bibfield{author}{\bibinfo{person}{Aditya Bhattacharya} {and} \bibinfo{person}{Katrien Verbert}.} \bibinfo{year}{2024}\natexlab{}.
\newblock \showarticletitle{"How Good Is Your Explanation?": Towards a Standardised Evaluation Approach for Diverse XAI Methods on Multiple Dimensions of Explainability}. In \bibinfo{booktitle}{\emph{Adjunct Proceedings of the 32nd ACM Conference on User Modeling, Adaptation and Personalization}} (Cagliari, Italy) \emph{(\bibinfo{series}{UMAP Adjunct '24})}. \bibinfo{publisher}{Association for Computing Machinery}, \bibinfo{address}{New York, NY, USA}, \bibinfo{pages}{513–515}.
\newblock
\showISBNx{9798400704666}
\urldef\tempurl%
\url{https://doi.org/10.1145/3631700.3664911}
\showDOI{\tempurl}


\bibitem[bin Ahmad and Takemoto(2024)]%
        {ahmad2024largescalemoralmachineexperiment}
\bibfield{author}{\bibinfo{person}{Muhammad Shahrul~Zaim bin Ahmad} {and} \bibinfo{person}{Kazuhiro Takemoto}.} \bibinfo{year}{2024}\natexlab{}.
\newblock \bibinfo{title}{Large-scale moral machine experiment on large language models}.
\newblock
\newblock
\showeprint[arxiv]{2411.06790}~[cs.CY]
\urldef\tempurl%
\url{https://arxiv.org/abs/2411.06790}
\showURL{%
\tempurl}


\bibitem[Bo et~al\mbox{.}(2025)]%
        {bo2025relyrelyevaluatinginterventions}
\bibfield{author}{\bibinfo{person}{Jessica~Y. Bo}, \bibinfo{person}{Sophia Wan}, {and} \bibinfo{person}{Ashton Anderson}.} \bibinfo{year}{2025}\natexlab{}.
\newblock \bibinfo{title}{To Rely or Not to Rely? Evaluating Interventions for Appropriate Reliance on Large Language Models}.
\newblock
\newblock
\showeprint[arxiv]{2412.15584}~[cs.HC]
\urldef\tempurl%
\url{https://arxiv.org/abs/2412.15584}
\showURL{%
\tempurl}


\bibitem[Braun and Clarke(2012)]%
        {BraunClarkTA}
\bibfield{author}{\bibinfo{person}{Virginia Braun} {and} \bibinfo{person}{Victoria Clarke}.} \bibinfo{year}{2012}\natexlab{}.
\newblock \showarticletitle{Thematic Analysis}.
\newblock In \bibinfo{booktitle}{\emph{{{APA}} Handbook of Research Methods in Psychology, {{Vol}} 2: {{Research}} Designs: {{Quantitative}}, Qualitative, Neuropsychological, and Biological}}. \bibinfo{publisher}{{American Psychological Association}}, \bibinfo{address}{{Washington, DC, US}}, \bibinfo{pages}{57--71}.
\newblock
\showISBNx{978-1-4338-1005-3}
\urldef\tempurl%
\url{https://doi.org/10.1037/13620-004}
\showDOI{\tempurl}


\bibitem[Caetano et~al\mbox{.}(2025)]%
        {caetano2025agenticworkflowsconversationalhumanai}
\bibfield{author}{\bibinfo{person}{Arthur Caetano}, \bibinfo{person}{Kavya Verma}, \bibinfo{person}{Atieh Taheri}, \bibinfo{person}{Radha Kumaran}, \bibinfo{person}{Zichen Chen}, \bibinfo{person}{Jiaao Chen}, \bibinfo{person}{Tobias Höllerer}, {and} \bibinfo{person}{Misha Sra}.} \bibinfo{year}{2025}\natexlab{}.
\newblock \bibinfo{title}{Agentic Workflows for Conversational Human-AI Interaction Design}.
\newblock
\newblock
\showeprint[arxiv]{2501.18002}~[cs.HC]
\urldef\tempurl%
\url{https://arxiv.org/abs/2501.18002}
\showURL{%
\tempurl}


\bibitem[Chen et~al\mbox{.}(2023)]%
        {chen2023understandingrolehumanintuition}
\bibfield{author}{\bibinfo{person}{Valerie Chen}, \bibinfo{person}{Q.~Vera Liao}, \bibinfo{person}{Jennifer~Wortman Vaughan}, {and} \bibinfo{person}{Gagan Bansal}.} \bibinfo{year}{2023}\natexlab{}.
\newblock \bibinfo{title}{Understanding the Role of Human Intuition on Reliance in Human-AI Decision-Making with Explanations}.
\newblock
\newblock
\showeprint[arxiv]{2301.07255}~[cs.HC]
\urldef\tempurl%
\url{https://arxiv.org/abs/2301.07255}
\showURL{%
\tempurl}


\bibitem[Christakopoulou et~al\mbox{.}(2024)]%
        {christakopoulou2024agentsthinkingfastslow}
\bibfield{author}{\bibinfo{person}{Konstantina Christakopoulou}, \bibinfo{person}{Shibl Mourad}, {and} \bibinfo{person}{Maja Matarić}.} \bibinfo{year}{2024}\natexlab{}.
\newblock \bibinfo{title}{Agents Thinking Fast and Slow: A Talker-Reasoner Architecture}.
\newblock
\newblock
\showeprint[arxiv]{2410.08328}~[cs.AI]
\urldef\tempurl%
\url{https://arxiv.org/abs/2410.08328}
\showURL{%
\tempurl}


\bibitem[Diamond and Banerjee(2024)]%
        {diamond2024ethicalconsiderationsgenerativeagents}
\bibfield{author}{\bibinfo{person}{N'yoma Diamond} {and} \bibinfo{person}{Soumya Banerjee}.} \bibinfo{year}{2024}\natexlab{}.
\newblock \bibinfo{title}{On the Ethical Considerations of Generative Agents}.
\newblock
\newblock
\showeprint[arxiv]{2411.19211}~[cs.CY]
\urldef\tempurl%
\url{https://arxiv.org/abs/2411.19211}
\showURL{%
\tempurl}


\bibitem[Fourney et~al\mbox{.}(2024)]%
        {fourney2024magenticonegeneralistmultiagentsolving}
\bibfield{author}{\bibinfo{person}{Adam Fourney}, \bibinfo{person}{Gagan Bansal}, \bibinfo{person}{Hussein Mozannar}, \bibinfo{person}{Cheng Tan}, \bibinfo{person}{Eduardo Salinas}, \bibinfo{person}{Erkang}, \bibinfo{person}{Zhu}, \bibinfo{person}{Friederike Niedtner}, \bibinfo{person}{Grace Proebsting}, \bibinfo{person}{Griffin Bassman}, \bibinfo{person}{Jack Gerrits}, \bibinfo{person}{Jacob Alber}, \bibinfo{person}{Peter Chang}, \bibinfo{person}{Ricky Loynd}, \bibinfo{person}{Robert West}, \bibinfo{person}{Victor Dibia}, \bibinfo{person}{Ahmed Awadallah}, \bibinfo{person}{Ece Kamar}, \bibinfo{person}{Rafah Hosn}, {and} \bibinfo{person}{Saleema Amershi}.} \bibinfo{year}{2024}\natexlab{}.
\newblock \bibinfo{title}{Magentic-One: A Generalist Multi-Agent System for Solving Complex Tasks}.
\newblock
\newblock
\showeprint[arxiv]{2411.04468}~[cs.AI]
\urldef\tempurl%
\url{https://arxiv.org/abs/2411.04468}
\showURL{%
\tempurl}


\bibitem[Fredes and Vitria(2024)]%
        {fredes2024usingllmsexplainingsets}
\bibfield{author}{\bibinfo{person}{Arturo Fredes} {and} \bibinfo{person}{Jordi Vitria}.} \bibinfo{year}{2024}\natexlab{}.
\newblock \bibinfo{title}{Using LLMs for Explaining Sets of Counterfactual Examples to Final Users}.
\newblock
\newblock
\showeprint[arxiv]{2408.15133}~[cs.LG]
\urldef\tempurl%
\url{https://arxiv.org/abs/2408.15133}
\showURL{%
\tempurl}


\bibitem[Freed and Safari(2021)]%
        {freed2021conversational}
\bibfield{author}{\bibinfo{person}{A. Freed} {and} \bibinfo{person}{an~O'Reilly Media~Company Safari}.} \bibinfo{year}{2021}\natexlab{}.
\newblock \bibinfo{booktitle}{\emph{Conversational AI}}.
\newblock \bibinfo{publisher}{Manning Publications}.
\newblock
\urldef\tempurl%
\url{https://books.google.be/books?id=wtKAzwEACAAJ}
\showURL{%
\tempurl}


\bibitem[Gan et~al\mbox{.}(2024)]%
        {gan2024applicationllmagentsrecruitment}
\bibfield{author}{\bibinfo{person}{Chengguang Gan}, \bibinfo{person}{Qinghao Zhang}, {and} \bibinfo{person}{Tatsunori Mori}.} \bibinfo{year}{2024}\natexlab{}.
\newblock \bibinfo{title}{Application of LLM Agents in Recruitment: A Novel Framework for Resume Screening}.
\newblock
\newblock
\showeprint[arxiv]{2401.08315}~[cs.CL]
\urldef\tempurl%
\url{https://arxiv.org/abs/2401.08315}
\showURL{%
\tempurl}


\bibitem[Gao et~al\mbox{.}(2024)]%
        {gao2024retrievalaugmentedgenerationlargelanguage}
\bibfield{author}{\bibinfo{person}{Yunfan Gao}, \bibinfo{person}{Yun Xiong}, \bibinfo{person}{Xinyu Gao}, \bibinfo{person}{Kangxiang Jia}, \bibinfo{person}{Jinliu Pan}, \bibinfo{person}{Yuxi Bi}, \bibinfo{person}{Yi Dai}, \bibinfo{person}{Jiawei Sun}, \bibinfo{person}{Meng Wang}, {and} \bibinfo{person}{Haofen Wang}.} \bibinfo{year}{2024}\natexlab{}.
\newblock \bibinfo{title}{Retrieval-Augmented Generation for Large Language Models: A Survey}.
\newblock
\newblock
\showeprint[arxiv]{2312.10997}~[cs.CL]
\urldef\tempurl%
\url{https://arxiv.org/abs/2312.10997}
\showURL{%
\tempurl}


\bibitem[Guo et~al\mbox{.}(2024)]%
        {guo2024largelanguagemodelbased}
\bibfield{author}{\bibinfo{person}{Taicheng Guo}, \bibinfo{person}{Xiuying Chen}, \bibinfo{person}{Yaqi Wang}, \bibinfo{person}{Ruidi Chang}, \bibinfo{person}{Shichao Pei}, \bibinfo{person}{Nitesh~V. Chawla}, \bibinfo{person}{Olaf Wiest}, {and} \bibinfo{person}{Xiangliang Zhang}.} \bibinfo{year}{2024}\natexlab{}.
\newblock \bibinfo{title}{Large Language Model based Multi-Agents: A Survey of Progress and Challenges}.
\newblock
\newblock
\showeprint[arxiv]{2402.01680}~[cs.CL]
\urldef\tempurl%
\url{https://arxiv.org/abs/2402.01680}
\showURL{%
\tempurl}


\bibitem[Han et~al\mbox{.}(2024)]%
        {han2024llmmultiagentsystemschallenges}
\bibfield{author}{\bibinfo{person}{Shanshan Han}, \bibinfo{person}{Qifan Zhang}, \bibinfo{person}{Yuhang Yao}, \bibinfo{person}{Weizhao Jin}, \bibinfo{person}{Zhaozhuo Xu}, {and} \bibinfo{person}{Chaoyang He}.} \bibinfo{year}{2024}\natexlab{}.
\newblock \bibinfo{title}{LLM Multi-Agent Systems: Challenges and Open Problems}.
\newblock
\newblock
\showeprint[arxiv]{2402.03578}~[cs.MA]
\urldef\tempurl%
\url{https://arxiv.org/abs/2402.03578}
\showURL{%
\tempurl}


\bibitem[Hart and Staveland(1988)]%
        {HART1988139}
\bibfield{author}{\bibinfo{person}{Sandra~G. Hart} {and} \bibinfo{person}{Lowell~E. Staveland}.} \bibinfo{year}{1988}\natexlab{}.
\newblock \showarticletitle{Development of NASA-TLX (Task Load Index): Results of Empirical and Theoretical Research}.
\newblock In \bibinfo{booktitle}{\emph{Human Mental Workload}}, \bibfield{editor}{\bibinfo{person}{Peter~A. Hancock} {and} \bibinfo{person}{Najmedin Meshkati}} (Eds.). \bibinfo{series}{Advances in Psychology}, Vol.~\bibinfo{volume}{52}. \bibinfo{publisher}{North-Holland}, \bibinfo{pages}{139--183}.
\newblock
\showISSN{0166-4115}
\urldef\tempurl%
\url{https://doi.org/10.1016/S0166-4115(08)62386-9}
\showDOI{\tempurl}


\bibitem[He et~al\mbox{.}(2025)]%
        {he2025plan}
\bibfield{author}{\bibinfo{person}{Gaole He}, \bibinfo{person}{Gianluca Demartini}, {and} \bibinfo{person}{Ujwal Gadiraju}.} \bibinfo{year}{2025}\natexlab{}.
\newblock \showarticletitle{Plan-Then-Execute: An Empirical Study of User Trust and Team Performance When Using LLM Agents As A Daily Assistant}. In \bibinfo{booktitle}{\emph{CHI Conference on Human Factors in Computing Systems (CHI '25)}} (Yokohama, Japan) \emph{(\bibinfo{series}{CHI '25})}. \bibinfo{publisher}{ACM}, \bibinfo{address}{New York, NY, USA}, \bibinfo{pages}{22}.
\newblock
\urldef\tempurl%
\url{https://doi.org/10.1145/3706598.3713218}
\showDOI{\tempurl}


\bibitem[Hou et~al\mbox{.}(2024)]%
        {LLMMemory2024}
\bibfield{author}{\bibinfo{person}{Yuki Hou}, \bibinfo{person}{Haruki Tamoto}, {and} \bibinfo{person}{Homei Miyashita}.} \bibinfo{year}{2024}\natexlab{}.
\newblock \showarticletitle{"My agent understands me better": Integrating Dynamic Human-like Memory Recall and Consolidation in LLM-Based Agents}. In \bibinfo{booktitle}{\emph{Extended Abstracts of the CHI Conference on Human Factors in Computing Systems}} (Honolulu, HI, USA) \emph{(\bibinfo{series}{CHI EA '24})}. \bibinfo{publisher}{Association for Computing Machinery}, \bibinfo{address}{New York, NY, USA}, Article \bibinfo{articleno}{7}, \bibinfo{numpages}{7}~pages.
\newblock
\showISBNx{9798400703317}
\urldef\tempurl%
\url{https://doi.org/10.1145/3613905.3650839}
\showDOI{\tempurl}


\bibitem[Huang et~al\mbox{.}(2023)]%
        {huang2023surveyhallucinationlargelanguage}
\bibfield{author}{\bibinfo{person}{Lei Huang}, \bibinfo{person}{Weijiang Yu}, \bibinfo{person}{Weitao Ma}, \bibinfo{person}{Weihong Zhong}, \bibinfo{person}{Zhangyin Feng}, \bibinfo{person}{Haotian Wang}, \bibinfo{person}{Qianglong Chen}, \bibinfo{person}{Weihua Peng}, \bibinfo{person}{Xiaocheng Feng}, \bibinfo{person}{Bing Qin}, {and} \bibinfo{person}{Ting Liu}.} \bibinfo{year}{2023}\natexlab{}.
\newblock \bibinfo{title}{A Survey on Hallucination in Large Language Models: Principles, Taxonomy, Challenges, and Open Questions}.
\newblock
\newblock
\showeprint[arxiv]{2311.05232}~[cs.CL]
\urldef\tempurl%
\url{https://arxiv.org/abs/2311.05232}
\showURL{%
\tempurl}


\bibitem[Hunkenschroer and Luetge(2022)]%
        {hunkenschroer2022ethics}
\bibfield{author}{\bibinfo{person}{Anna-Lena Hunkenschroer} {and} \bibinfo{person}{Christoph Luetge}.} \bibinfo{year}{2022}\natexlab{}.
\newblock \showarticletitle{Ethics of AI-Enabled Recruiting and Selection: A Review and Research Agenda}.
\newblock \bibinfo{journal}{\emph{Journal of Business Ethics}}  \bibinfo{volume}{178} (\bibinfo{year}{2022}), \bibinfo{pages}{977--1007}.
\newblock
\urldef\tempurl%
\url{https://doi.org/10.1007/s10551-022-05049-6}
\showDOI{\tempurl}


\bibitem[Jian et~al\mbox{.}(2000)]%
        {Jian2020_trust_scale}
\bibfield{author}{\bibinfo{person}{Jiun-Yin Jian}, \bibinfo{person}{Ann Bisantz}, {and} \bibinfo{person}{Colin Drury}.} \bibinfo{year}{2000}\natexlab{}.
\newblock \showarticletitle{Foundations for an Empirically Determined Scale of Trust in Automated Systems}.
\newblock \bibinfo{journal}{\emph{International Journal of Cognitive Ergonomics}}  \bibinfo{volume}{4} (\bibinfo{date}{03} \bibinfo{year}{2000}), \bibinfo{pages}{53--71}.
\newblock
\urldef\tempurl%
\url{https://doi.org/10.1207/S15327566IJCE0401_04}
\showDOI{\tempurl}


\bibitem[Kim et~al\mbox{.}(2025)]%
        {kim2025fostering}
\bibfield{author}{\bibinfo{person}{Sunnie S.~Y. Kim}, \bibinfo{person}{Jennifer~Wortman Vaughan}, \bibinfo{person}{Q.~Vera Liao}, \bibinfo{person}{Tania Lombrozo}, {and} \bibinfo{person}{Olga Russakovsky}.} \bibinfo{year}{2025}\natexlab{}.
\newblock \showarticletitle{Fostering Appropriate Reliance on Large Language Models: The Role of Explanations, Sources, and Inconsistencies}. In \bibinfo{booktitle}{\emph{Proceedings of the CHI Conference on Human Factors in Computing Systems (CHI '25)}}. \bibinfo{publisher}{ACM}, \bibinfo{address}{Yokohama, Japan}, \bibinfo{pages}{1--26}.
\newblock
\urldef\tempurl%
\url{https://doi.org/10.1145/3706598.3714020}
\showDOI{\tempurl}


\bibitem[Kunz and Kuhlmann(2024)]%
        {kunz-kuhlmann-2024-properties}
\bibfield{author}{\bibinfo{person}{Jenny Kunz} {and} \bibinfo{person}{Marco Kuhlmann}.} \bibinfo{year}{2024}\natexlab{}.
\newblock \showarticletitle{Properties and Challenges of {LLM}-Generated Explanations}. In \bibinfo{booktitle}{\emph{Proceedings of the Third Workshop on Bridging Human--Computer Interaction and Natural Language Processing}}, \bibfield{editor}{\bibinfo{person}{Su~Lin Blodgett}, \bibinfo{person}{Amanda Cercas~Curry}, \bibinfo{person}{Sunipa Dev}, \bibinfo{person}{Michael Madaio}, \bibinfo{person}{Ani Nenkova}, \bibinfo{person}{Diyi Yang}, {and} \bibinfo{person}{Ziang Xiao}} (Eds.). \bibinfo{publisher}{Association for Computational Linguistics}, \bibinfo{address}{Mexico City, Mexico}, \bibinfo{pages}{13--27}.
\newblock
\urldef\tempurl%
\url{https://doi.org/10.18653/v1/2024.hcinlp-1.2}
\showDOI{\tempurl}


\bibitem[Köchling and Wehner(2020)]%
        {Kochling2020}
\bibfield{author}{\bibinfo{person}{Andreas Köchling} {and} \bibinfo{person}{Marius~C. Wehner}.} \bibinfo{year}{2020}\natexlab{}.
\newblock \showarticletitle{Discriminated by an algorithm: a systematic review of discrimination and fairness by algorithmic decision-making in the context of HR recruitment and HR development}.
\newblock \bibinfo{journal}{\emph{Business Research}}  \bibinfo{volume}{13} (\bibinfo{year}{2020}), \bibinfo{pages}{795--848}.
\newblock
\urldef\tempurl%
\url{https://doi.org/10.1007/s40685-020-00134-w}
\showDOI{\tempurl}


\bibitem[Lakkaraju et~al\mbox{.}(2022)]%
        {lakkaraju2022rethinking}
\bibfield{author}{\bibinfo{person}{Himabindu Lakkaraju}, \bibinfo{person}{Dylan Slack}, \bibinfo{person}{Yuxin Chen}, \bibinfo{person}{Chenhao Tan}, {and} \bibinfo{person}{Sameer Singh}.} \bibinfo{year}{2022}\natexlab{}.
\newblock \bibinfo{title}{Rethinking Explainability as a Dialogue: A Practitioner's Perspective}.
\newblock
\newblock
\showeprint[arxiv]{2202.01875}~[cs.LG]


\bibitem[LangChain(2025)]%
        {langchain_online}
\bibfield{author}{\bibinfo{person}{LangChain}.} \bibinfo{year}{2025}\natexlab{}.
\newblock \bibinfo{booktitle}{}.
\newblock LangChain.
\newblock
\urldef\tempurl%
\url{https://www.langchain.com}
\showURL{%
\tempurl}
\newblock
\shownote{Accessed: 2025-01-21}.


\bibitem[Li et~al\mbox{.}(2024a)]%
        {BrennaLi2024}
\bibfield{author}{\bibinfo{person}{Brenna Li}, \bibinfo{person}{Ofek Gross}, \bibinfo{person}{Noah Crampton}, \bibinfo{person}{Mamta Kapoor}, \bibinfo{person}{Saba Tauseef}, \bibinfo{person}{Mohit Jain}, \bibinfo{person}{Khai~N. Truong}, {and} \bibinfo{person}{Alex Mariakakis}.} \bibinfo{year}{2024}\natexlab{a}.
\newblock \showarticletitle{Beyond the Waiting Room: Patient's Perspectives on the Conversational Nuances of Pre-Consultation Chatbots}. In \bibinfo{booktitle}{\emph{Proceedings of the 2024 CHI Conference on Human Factors in Computing Systems}} (Honolulu, HI, USA) \emph{(\bibinfo{series}{CHI '24})}. \bibinfo{publisher}{Association for Computing Machinery}, \bibinfo{address}{New York, NY, USA}, Article \bibinfo{articleno}{438}, \bibinfo{numpages}{24}~pages.
\newblock
\showISBNx{9798400703300}
\urldef\tempurl%
\url{https://doi.org/10.1145/3613904.3641913}
\showDOI{\tempurl}


\bibitem[Li et~al\mbox{.}(2024b)]%
        {li2024MASarchitetcuresurvey}
\bibfield{author}{\bibinfo{person}{X. Li}, \bibinfo{person}{S. Wang}, \bibinfo{person}{S. Zeng}, {et~al\mbox{.}}} \bibinfo{year}{2024}\natexlab{b}.
\newblock \showarticletitle{A survey on LLM-based multi-agent systems: workflow, infrastructure, and challenges}.
\newblock \bibinfo{journal}{\emph{Vicinagearth}} \bibinfo{volume}{1}, \bibinfo{number}{9} (\bibinfo{year}{2024}).
\newblock
\urldef\tempurl%
\url{https://doi.org/10.1007/s44336-024-00009-2}
\showDOI{\tempurl}


\bibitem[Liu et~al\mbox{.}(2024)]%
        {liu2024promptinjectionattackllmintegrated}
\bibfield{author}{\bibinfo{person}{Yi Liu}, \bibinfo{person}{Gelei Deng}, \bibinfo{person}{Yuekang Li}, \bibinfo{person}{Kailong Wang}, \bibinfo{person}{Zihao Wang}, \bibinfo{person}{Xiaofeng Wang}, \bibinfo{person}{Tianwei Zhang}, \bibinfo{person}{Yepang Liu}, \bibinfo{person}{Haoyu Wang}, \bibinfo{person}{Yan Zheng}, {and} \bibinfo{person}{Yang Liu}.} \bibinfo{year}{2024}\natexlab{}.
\newblock \bibinfo{title}{Prompt Injection attack against LLM-integrated Applications}.
\newblock
\newblock
\showeprint[arxiv]{2306.05499}~[cs.CR]
\urldef\tempurl%
\url{https://arxiv.org/abs/2306.05499}
\showURL{%
\tempurl}


\bibitem[Luo et~al\mbox{.}(2023)]%
        {luo2023providingpersonalizedexplanationsconversational}
\bibfield{author}{\bibinfo{person}{Jieting Luo}, \bibinfo{person}{Thomas Studer}, {and} \bibinfo{person}{Mehdi Dastani}.} \bibinfo{year}{2023}\natexlab{}.
\newblock \bibinfo{title}{Providing personalized Explanations: a Conversational Approach}.
\newblock
\newblock
\showeprint[arxiv]{2307.11452}~[cs.MA]
\urldef\tempurl%
\url{https://arxiv.org/abs/2307.11452}
\showURL{%
\tempurl}


\bibitem[Luo et~al\mbox{.}(2025)]%
        {Luo2025EARNFairness}
\bibfield{author}{\bibinfo{person}{Lin Luo}, \bibinfo{person}{Yuri Nakao}, \bibinfo{person}{Mathieu Chollet}, \bibinfo{person}{Hiroya Inakoshi}, {and} \bibinfo{person}{Simone Stumpf}.} \bibinfo{year}{2025}\natexlab{}.
\newblock \showarticletitle{{EARN Fairness: Explaining, Asking, Reviewing, and Negotiating Artificial Intelligence Fairness Metrics Among Stakeholders}}.
\newblock \bibinfo{journal}{\emph{Proceedings of the ACM on Human-Computer Interaction}} \bibinfo{volume}{9}, \bibinfo{number}{2} (\bibinfo{date}{April} \bibinfo{year}{2025}), \bibinfo{pages}{37}.
\newblock
\urldef\tempurl%
\url{https://doi.org/10.1145/3710908}
\showDOI{\tempurl}


\bibitem[Mariani et~al\mbox{.}(2023)]%
        {Marini2023CA}
\bibfield{author}{\bibinfo{person}{Marcello~M. Mariani}, \bibinfo{person}{Novin Hashemi}, {and} \bibinfo{person}{Jochen Wirtz}.} \bibinfo{year}{2023}\natexlab{}.
\newblock \showarticletitle{Artificial intelligence empowered conversational agents: A systematic literature review and research agenda}.
\newblock \bibinfo{journal}{\emph{Journal of Business Research}}  \bibinfo{volume}{161} (\bibinfo{year}{2023}), \bibinfo{pages}{113838}.
\newblock
\showISSN{0148-2963}
\urldef\tempurl%
\url{https://doi.org/10.1016/j.jbusres.2023.113838}
\showDOI{\tempurl}


\bibitem[McCrum-Gardner(2008)]%
        {mccrum-gardner_which_2008}
\bibfield{author}{\bibinfo{person}{Evie McCrum-Gardner}.} \bibinfo{year}{2008}\natexlab{}.
\newblock \showarticletitle{Which is the correct statistical test to use?}
\newblock \bibinfo{journal}{\emph{British Journal of Oral and Maxillofacial Surgery}} \bibinfo{volume}{46}, \bibinfo{number}{1} (\bibinfo{date}{Jan.} \bibinfo{year}{2008}), \bibinfo{pages}{38--41}.
\newblock
\showISSN{02664356}
\urldef\tempurl%
\url{https://doi.org/10.1016/j.bjoms.2007.09.002}
\showDOI{\tempurl}


\bibitem[Mehrotra et~al\mbox{.}(2024)]%
        {MehrotraTrust2024}
\bibfield{author}{\bibinfo{person}{Siddharth Mehrotra}, \bibinfo{person}{Chadha Degachi}, \bibinfo{person}{Oleksandra Vereschak}, \bibinfo{person}{Catholijn~M. Jonker}, {and} \bibinfo{person}{Myrthe~L. Tielman}.} \bibinfo{year}{2024}\natexlab{}.
\newblock \showarticletitle{A Systematic Review on Fostering Appropriate Trust in Human-AI Interaction: Trends, Opportunities and Challenges}.
\newblock \bibinfo{journal}{\emph{ACM J. Responsib. Comput.}} \bibinfo{volume}{1}, \bibinfo{number}{4}, Article \bibinfo{articleno}{26} (\bibinfo{date}{Nov.} \bibinfo{year}{2024}), \bibinfo{numpages}{45}~pages.
\newblock
\urldef\tempurl%
\url{https://doi.org/10.1145/3696449}
\showDOI{\tempurl}


\bibitem[Miller(2017)]%
        {Miller2017}
\bibfield{author}{\bibinfo{person}{Tim Miller}.} \bibinfo{year}{2017}\natexlab{}.
\newblock \bibinfo{title}{Explanation in Artificial Intelligence: Insights from the Social Sciences}.
\newblock
\newblock
\urldef\tempurl%
\url{https://doi.org/10.48550/ARXIV.1706.07269}
\showDOI{\tempurl}


\bibitem[Nakao et~al\mbox{.}(2022)]%
        {Nakao2022}
\bibfield{author}{\bibinfo{person}{Yuri Nakao}, \bibinfo{person}{Simone Stumpf}, \bibinfo{person}{Subeida Ahmed}, \bibinfo{person}{Aisha Naseer}, {and} \bibinfo{person}{Lorenzo Strappelli}.} \bibinfo{year}{2022}\natexlab{}.
\newblock \showarticletitle{Toward Involving End-users in Interactive Human-in-the-loop AI Fairness}.
\newblock \bibinfo{journal}{\emph{ACM Trans. Interact. Intell. Syst.}} \bibinfo{volume}{12}, \bibinfo{number}{3}, Article \bibinfo{articleno}{18} (\bibinfo{date}{July} \bibinfo{year}{2022}), \bibinfo{numpages}{30}~pages.
\newblock
\showISSN{2160-6455}
\urldef\tempurl%
\url{https://doi.org/10.1145/3514258}
\showDOI{\tempurl}


\bibitem[OpenAI(2025)]%
        {openai_online}
\bibfield{author}{\bibinfo{person}{OpenAI}.} \bibinfo{year}{2025}\natexlab{}.
\newblock \bibinfo{booktitle}{}.
\newblock OpenAI.
\newblock
\urldef\tempurl%
\url{https://openai.com/index/openai-api/}
\showURL{%
\tempurl}
\newblock
\shownote{Accessed: 2025-01-21}.


\bibitem[{Oxford University Press}(2014)]%
        {oxford_zerosum}
\bibfield{author}{\bibinfo{person}{{Oxford University Press}}.} \bibinfo{year}{2014}\natexlab{}.
\newblock \bibinfo{booktitle}{\emph{Zero-sum game}}.
\newblock
\urldef\tempurl%
\url{https://www.oxfordreference.com/display/10.1093/acref/9780199670840.001.0001/acref-9780199670840-e-1491}
\showURL{%
\tempurl}
\newblock
\shownote{Accessed: 2025-03-29}.


\bibitem[Passi and Vorvoreanu(2022)]%
        {PassiVorvoreanu2022}
\bibfield{author}{\bibinfo{person}{Samir Passi} {and} \bibinfo{person}{Mihaela Vorvoreanu}.} \bibinfo{year}{2022}\natexlab{}.
\newblock \bibinfo{booktitle}{\emph{Overreliance on AI: Literature Review}}.
\newblock \bibinfo{type}{Microsoft Technical Report} MSR-TR-2022-12. \bibinfo{institution}{Microsoft Corporation}.
\newblock
\urldef\tempurl%
\url{https://www.microsoft.com/en-us/research/uploads/prod/2022/06/Aether-Overreliance-on-AI-Review-Final-6.21.22.pdf}
\showURL{%
\tempurl}


\bibitem[Rahn et~al\mbox{.}(2024)]%
        {rahn2024controllinglargelanguagemodel}
\bibfield{author}{\bibinfo{person}{Nate Rahn}, \bibinfo{person}{Pierluca D'Oro}, {and} \bibinfo{person}{Marc~G. Bellemare}.} \bibinfo{year}{2024}\natexlab{}.
\newblock \bibinfo{title}{Controlling Large Language Model Agents with Entropic Activation Steering}.
\newblock
\newblock
\showeprint[arxiv]{2406.00244}~[cs.CL]
\urldef\tempurl%
\url{https://arxiv.org/abs/2406.00244}
\showURL{%
\tempurl}


\bibitem[Ritter et~al\mbox{.}(2014)]%
        {UCD2014}
\bibfield{author}{\bibinfo{person}{Frank Ritter}, \bibinfo{person}{Gordon Baxter}, {and} \bibinfo{person}{Elizabeth Churchill}.} \bibinfo{year}{2014}\natexlab{}.
\newblock \bibinfo{booktitle}{\emph{User-Centered Systems Design: A Brief History}}.
\newblock \bibinfo{publisher}{Springer}, \bibinfo{pages}{33--54}.
\newblock
\showISBNx{978-1-4471-5133-3}
\urldef\tempurl%
\url{https://doi.org/10.1007/978-1-4471-5134-0_2}
\showDOI{\tempurl}


\bibitem[Safavi-Naini et~al\mbox{.}(2024)]%
        {safavinaini2024visionlanguagelargelanguagemodel}
\bibfield{author}{\bibinfo{person}{Seyed Amir~Ahmad Safavi-Naini}, \bibinfo{person}{Shuhaib Ali}, \bibinfo{person}{Omer Shahab}, \bibinfo{person}{Zahra Shahhoseini}, \bibinfo{person}{Thomas Savage}, \bibinfo{person}{Sara Rafiee}, \bibinfo{person}{Jamil~S Samaan}, \bibinfo{person}{Reem~Al Shabeeb}, \bibinfo{person}{Farah Ladak}, \bibinfo{person}{Jamie~O Yang}, \bibinfo{person}{Juan Echavarria}, \bibinfo{person}{Sumbal Babar}, \bibinfo{person}{Aasma Shaukat}, \bibinfo{person}{Samuel Margolis}, \bibinfo{person}{Nicholas~P Tatonetti}, \bibinfo{person}{Girish Nadkarni}, \bibinfo{person}{Bara~El Kurdi}, {and} \bibinfo{person}{Ali Soroush}.} \bibinfo{year}{2024}\natexlab{}.
\newblock \bibinfo{title}{Vision-Language and Large Language Model Performance in Gastroenterology: GPT, Claude, Llama, Phi, Mistral, Gemma, and Quantized Models}.
\newblock
\newblock
\showeprint[arxiv]{2409.00084}~[cs.CL]
\urldef\tempurl%
\url{https://arxiv.org/abs/2409.00084}
\showURL{%
\tempurl}


\bibitem[Schiller(2024)]%
        {schiller2024humanfactordetectingerrors}
\bibfield{author}{\bibinfo{person}{Christian~A. Schiller}.} \bibinfo{year}{2024}\natexlab{}.
\newblock \bibinfo{title}{The Human Factor in Detecting Errors of Large Language Models: A Systematic Literature Review and Future Research Directions}.
\newblock
\newblock
\showeprint[arxiv]{2403.09743}~[cs.CL]
\urldef\tempurl%
\url{https://arxiv.org/abs/2403.09743}
\showURL{%
\tempurl}


\bibitem[Sharma et~al\mbox{.}(2024)]%
        {sharma-etal-2024-investigating}
\bibfield{author}{\bibinfo{person}{Ashish Sharma}, \bibinfo{person}{Sudha Rao}, \bibinfo{person}{Chris Brockett}, \bibinfo{person}{Akanksha Malhotra}, \bibinfo{person}{Nebojsa Jojic}, {and} \bibinfo{person}{Bill Dolan}.} \bibinfo{year}{2024}\natexlab{}.
\newblock \showarticletitle{Investigating Agency of {LLM}s in Human-{AI} Collaboration Tasks}. In \bibinfo{booktitle}{\emph{Proceedings of the 18th Conference of the European Chapter of the Association for Computational Linguistics (Volume 1: Long Papers)}}, \bibfield{editor}{\bibinfo{person}{Yvette Graham} {and} \bibinfo{person}{Matthew Purver}} (Eds.). \bibinfo{publisher}{Association for Computational Linguistics}, \bibinfo{address}{St. Julian{'}s, Malta}, \bibinfo{pages}{1968--1987}.
\newblock
\urldef\tempurl%
\url{https://aclanthology.org/2024.eacl-long.119/}
\showURL{%
\tempurl}


\bibitem[Shen et~al\mbox{.}(2023)]%
        {shen2023convxaideliveringheterogeneousai}
\bibfield{author}{\bibinfo{person}{Hua Shen}, \bibinfo{person}{Chieh-Yang Huang}, \bibinfo{person}{Tongshuang Wu}, {and} \bibinfo{person}{Ting-Hao~'Kenneth' Huang}.} \bibinfo{year}{2023}\natexlab{}.
\newblock \bibinfo{title}{ConvXAI: Delivering Heterogeneous AI Explanations via Conversations to Support Human-AI Scientific Writing}.
\newblock
\newblock
\showeprint[arxiv]{2305.09770}~[cs.HC]
\urldef\tempurl%
\url{https://arxiv.org/abs/2305.09770}
\showURL{%
\tempurl}


\bibitem[Shoemaker et~al\mbox{.}(2014)]%
        {shoemaker2014pemat}
\bibfield{author}{\bibinfo{person}{S.~J. Shoemaker}, \bibinfo{person}{M.~S. Wolf}, {and} \bibinfo{person}{C. Brach}.} \bibinfo{year}{2014}\natexlab{}.
\newblock \showarticletitle{Development of the patient education materials assessment tool (PEMAT): a new measure of understandability and actionability for print and audiovisual patient information}.
\newblock \bibinfo{journal}{\emph{Patient Education and Counseling}} \bibinfo{volume}{96}, \bibinfo{number}{3} (\bibinfo{date}{Sep} \bibinfo{year}{2014}), \bibinfo{pages}{395--403}.
\newblock
\urldef\tempurl%
\url{https://doi.org/10.1016/j.pec.2014.05.027}
\showDOI{\tempurl}


\bibitem[Singh et~al\mbox{.}(2024)]%
        {singh2024actionabilityassessmenttoolexplainable}
\bibfield{author}{\bibinfo{person}{Ronal Singh}, \bibinfo{person}{Tim Miller}, \bibinfo{person}{Liz Sonenberg}, \bibinfo{person}{Eduardo Velloso}, \bibinfo{person}{Frank Vetere}, \bibinfo{person}{Piers Howe}, {and} \bibinfo{person}{Paul Dourish}.} \bibinfo{year}{2024}\natexlab{}.
\newblock \bibinfo{title}{An Actionability Assessment Tool for Explainable AI}.
\newblock
\newblock
\showeprint[arxiv]{2407.09516}~[cs.HC]
\urldef\tempurl%
\url{https://arxiv.org/abs/2407.09516}
\showURL{%
\tempurl}


\bibitem[Slack et~al\mbox{.}(2023)]%
        {Slack2023}
\bibfield{author}{\bibinfo{person}{Dylan Slack}, \bibinfo{person}{Satyapriya Krishna}, \bibinfo{person}{Himabindu Lakkaraju}, {and} \bibinfo{person}{Sameer Singh}.} \bibinfo{year}{2023}\natexlab{}.
\newblock \showarticletitle{Explaining machine learning models with interactive natural language conversations using TalkToModel}.
\newblock \bibinfo{journal}{\emph{Nature Machine Intelligence}} (\bibinfo{date}{27 Jul} \bibinfo{year}{2023}).
\newblock
\showISSN{2522-5839}
\urldef\tempurl%
\url{https://doi.org/10.1038/s42256-023-00692-8}
\showDOI{\tempurl}


\bibitem[Song et~al\mbox{.}(2025)]%
        {SteerLLM2025}
\bibfield{author}{\bibinfo{person}{Bingqing Song}, \bibinfo{person}{Boran Han}, \bibinfo{person}{Shuai Zhang}, \bibinfo{person}{Hao Wang}, \bibinfo{person}{Haoyang Fang}, \bibinfo{person}{Bonan Min}, \bibinfo{person}{Yuyang Wang}, {and} \bibinfo{person}{Mingyi Hong}.} \bibinfo{year}{2025}\natexlab{}.
\newblock \bibinfo{title}{Effectively Steer LLM To Follow Preference via Building Confident Directions}.
\newblock
\newblock
\urldef\tempurl%
\url{https://doi.org/10.48550/arXiv.2503.02989}
\showDOI{\tempurl}


\bibitem[Stone and Shiffman(2002)]%
        {stone2002capturing}
\bibfield{author}{\bibinfo{person}{Arthur~A. Stone} {and} \bibinfo{person}{Saul Shiffman}.} \bibinfo{year}{2002}\natexlab{}.
\newblock \showarticletitle{Capturing momentary, self-report data: A proposal for reporting guidelines}.
\newblock \bibinfo{journal}{\emph{Annals of Behavioral Medicine}} \bibinfo{volume}{24}, \bibinfo{number}{3} (\bibinfo{year}{2002}), \bibinfo{pages}{236--243}.
\newblock
\urldef\tempurl%
\url{https://doi.org/10.1207/S15324796ABM2403_09}
\showDOI{\tempurl}


\bibitem[Streamlit(2025)]%
        {streamlit}
\bibfield{author}{\bibinfo{person}{Streamlit}.} \bibinfo{year}{2025}\natexlab{}.
\newblock \bibinfo{booktitle}{}.
\newblock Streamlit.
\newblock
\urldef\tempurl%
\url{https://streamlit.io}
\showURL{%
\tempurl}
\newblock
\shownote{Accessed: 2025-01-21}.


\bibitem[Talbert(2017)]%
        {analysisparalysis}
\bibfield{author}{\bibinfo{person}{Bonnie Talbert}.} \bibinfo{year}{2017}\natexlab{}.
\newblock \showarticletitle{Overthinking and Other Minds: The Analysis Paralysis}.
\newblock \bibinfo{journal}{\emph{Social Epistemology}}  \bibinfo{volume}{31} (\bibinfo{date}{07} \bibinfo{year}{2017}), \bibinfo{pages}{1--12}.
\newblock
\urldef\tempurl%
\url{https://doi.org/10.1080/02691728.2017.1346933}
\showDOI{\tempurl}


\bibitem[Tjuatja et~al\mbox{.}(2024)]%
        {tjuatja-etal-2024-llms}
\bibfield{author}{\bibinfo{person}{Lindia Tjuatja}, \bibinfo{person}{Valerie Chen}, \bibinfo{person}{Tongshuang Wu}, \bibinfo{person}{Ameet Talwalkwar}, {and} \bibinfo{person}{Graham Neubig}.} \bibinfo{year}{2024}\natexlab{}.
\newblock \showarticletitle{Do {LLM}s Exhibit Human-like Response Biases? A Case Study in Survey Design}.
\newblock \bibinfo{journal}{\emph{Transactions of the Association for Computational Linguistics}}  \bibinfo{volume}{12} (\bibinfo{year}{2024}), \bibinfo{pages}{1011--1026}.
\newblock
\urldef\tempurl%
\url{https://doi.org/10.1162/tacl_a_00685}
\showDOI{\tempurl}


\bibitem[Tran et~al\mbox{.}(2025)]%
        {tran2025multiagentcollaborationmechanismssurvey}
\bibfield{author}{\bibinfo{person}{Khanh-Tung Tran}, \bibinfo{person}{Dung Dao}, \bibinfo{person}{Minh-Duong Nguyen}, \bibinfo{person}{Quoc-Viet Pham}, \bibinfo{person}{Barry O'Sullivan}, {and} \bibinfo{person}{Hoang~D. Nguyen}.} \bibinfo{year}{2025}\natexlab{}.
\newblock \bibinfo{title}{Multi-Agent Collaboration Mechanisms: A Survey of LLMs}.
\newblock
\newblock
\showeprint[arxiv]{2501.06322}~[cs.AI]
\urldef\tempurl%
\url{https://arxiv.org/abs/2501.06322}
\showURL{%
\tempurl}


\bibitem[Wei et~al\mbox{.}(2022)]%
        {CoTPrompt2022}
\bibfield{author}{\bibinfo{person}{Jason Wei}, \bibinfo{person}{Xuezhi Wang}, \bibinfo{person}{Dale Schuurmans}, \bibinfo{person}{Maarten Bosma}, \bibinfo{person}{Brian Ichter}, \bibinfo{person}{Fei Xia}, \bibinfo{person}{Ed~H. Chi}, \bibinfo{person}{Quoc~V. Le}, {and} \bibinfo{person}{Denny Zhou}.} \bibinfo{year}{2022}\natexlab{}.
\newblock \showarticletitle{Chain-of-thought prompting elicits reasoning in large language models}. In \bibinfo{booktitle}{\emph{Proceedings of the 36th International Conference on Neural Information Processing Systems}} (New Orleans, LA, USA) \emph{(\bibinfo{series}{NIPS '22})}. \bibinfo{publisher}{Curran Associates Inc.}, \bibinfo{address}{Red Hook, NY, USA}, Article \bibinfo{articleno}{1800}, \bibinfo{numpages}{14}~pages.
\newblock
\showISBNx{9781713871088}


\bibitem[Weisz et~al\mbox{.}(2024)]%
        {Weisz2024DesignApplications}
\bibfield{author}{\bibinfo{person}{Justin~D. Weisz}, \bibinfo{person}{Jessica He}, \bibinfo{person}{Michael Muller}, \bibinfo{person}{Gabriela Hoefer}, \bibinfo{person}{Rachel Miles}, {and} \bibinfo{person}{Werner Geyer}.} \bibinfo{year}{2024}\natexlab{}.
\newblock \showarticletitle{{Design Principles for Generative AI Applications}}. In \bibinfo{booktitle}{\emph{Proceedings of the CHI Conference on Human Factors in Computing Systems}}. \bibinfo{publisher}{ACM}, \bibinfo{address}{New York, NY, USA}, \bibinfo{pages}{1--22}.
\newblock
\showISBNx{9798400703300}
\urldef\tempurl%
\url{https://doi.org/10.1145/3613904.3642466}
\showDOI{\tempurl}


\bibitem[Xie et~al\mbox{.}(2024)]%
        {WaitGPT2024}
\bibfield{author}{\bibinfo{person}{Liwenhan Xie}, \bibinfo{person}{Chengbo Zheng}, \bibinfo{person}{Haijun Xia}, \bibinfo{person}{Huamin Qu}, {and} \bibinfo{person}{Chen Zhu-Tian}.} \bibinfo{year}{2024}\natexlab{}.
\newblock \bibinfo{title}{WaitGPT: Monitoring and Steering Conversational LLM Agent in Data Analysis with On-the-Fly Code Visualization}.
\newblock
\newblock
\urldef\tempurl%
\url{https://doi.org/10.48550/arXiv.2408.01703}
\showDOI{\tempurl}


\bibitem[Xu et~al\mbox{.}(2024)]%
        {xu2024hallucinationinevitableinnatelimitation}
\bibfield{author}{\bibinfo{person}{Ziwei Xu}, \bibinfo{person}{Sanjay Jain}, {and} \bibinfo{person}{Mohan Kankanhalli}.} \bibinfo{year}{2024}\natexlab{}.
\newblock \bibinfo{title}{Hallucination is Inevitable: An Innate Limitation of Large Language Models}.
\newblock
\newblock
\showeprint[arxiv]{2401.11817}~[cs.CL]
\urldef\tempurl%
\url{https://arxiv.org/abs/2401.11817}
\showURL{%
\tempurl}


\bibitem[Yang and Aurisicchio(2021)]%
        {yangdesigningchatbot}
\bibfield{author}{\bibinfo{person}{Xi Yang} {and} \bibinfo{person}{Marco Aurisicchio}.} \bibinfo{year}{2021}\natexlab{}.
\newblock \showarticletitle{Designing Conversational Agents: A Self-Determination Theory Approach}. In \bibinfo{booktitle}{\emph{Proceedings of the 2021 CHI Conference on Human Factors in Computing Systems}} (Yokohama, Japan) \emph{(\bibinfo{series}{CHI '21})}. \bibinfo{publisher}{Association for Computing Machinery}, \bibinfo{address}{New York, NY, USA}, Article \bibinfo{articleno}{256}, \bibinfo{numpages}{16}~pages.
\newblock
\showISBNx{9781450380966}
\urldef\tempurl%
\url{https://doi.org/10.1145/3411764.3445445}
\showDOI{\tempurl}


\bibitem[Yao et~al\mbox{.}(2023)]%
        {yao2023react}
\bibfield{author}{\bibinfo{person}{Shunyu Yao}, \bibinfo{person}{Jeffrey Zhao}, \bibinfo{person}{Dian Yu}, \bibinfo{person}{Nan Du}, \bibinfo{person}{Izhak Shafran}, \bibinfo{person}{Karthik Narasimhan}, {and} \bibinfo{person}{Yuan Cao}.} \bibinfo{year}{2023}\natexlab{}.
\newblock \showarticletitle{{ReAct}: Synergizing Reasoning and Acting in Language Models}. In \bibinfo{booktitle}{\emph{International Conference on Learning Representations (ICLR)}}.
\newblock


\bibitem[Zhai et~al\mbox{.}(2024)]%
        {zhai2024effects}
\bibfield{author}{\bibinfo{person}{Cheng Zhai}, \bibinfo{person}{Syamsul Wibowo}, {and} \bibinfo{person}{L.D. Li}.} \bibinfo{year}{2024}\natexlab{}.
\newblock \showarticletitle{The effects of over-reliance on AI dialogue systems on students' cognitive abilities: a systematic review}.
\newblock \bibinfo{journal}{\emph{Smart Learning Environments}}  \bibinfo{volume}{11} (\bibinfo{year}{2024}).
\newblock
\urldef\tempurl%
\url{https://doi.org/10.1186/s40561-024-00316-7}
\showDOI{\tempurl}


\bibitem[Zhang et~al\mbox{.}(2024b)]%
        {zhang2024iaskfollowupquestion}
\bibfield{author}{\bibinfo{person}{Tong Zhang}, \bibinfo{person}{X.~Jessie Yang}, {and} \bibinfo{person}{Boyang Li}.} \bibinfo{year}{2024}\natexlab{b}.
\newblock \bibinfo{title}{May I Ask a Follow-up Question? Understanding the Benefits of Conversations in Neural Network Explainability}.
\newblock
\newblock
\showeprint[arxiv]{2309.13965}~[cs.HC]
\urldef\tempurl%
\url{https://arxiv.org/abs/2309.13965}
\showURL{%
\tempurl}


\bibitem[Zhang et~al\mbox{.}(2024a)]%
        {zhang2024surveymemorymechanismlarge}
\bibfield{author}{\bibinfo{person}{Zeyu Zhang}, \bibinfo{person}{Xiaohe Bo}, \bibinfo{person}{Chen Ma}, \bibinfo{person}{Rui Li}, \bibinfo{person}{Xu Chen}, \bibinfo{person}{Quanyu Dai}, \bibinfo{person}{Jieming Zhu}, \bibinfo{person}{Zhenhua Dong}, {and} \bibinfo{person}{Ji-Rong Wen}.} \bibinfo{year}{2024}\natexlab{a}.
\newblock \bibinfo{title}{A Survey on the Memory Mechanism of Large Language Model based Agents}.
\newblock
\newblock
\showeprint[arxiv]{2404.13501}~[cs.AI]
\urldef\tempurl%
\url{https://arxiv.org/abs/2404.13501}
\showURL{%
\tempurl}


\bibitem[Zheng et~al\mbox{.}(2023)]%
        {zheng2023judgingllmasajudgemtbenchchatbot}
\bibfield{author}{\bibinfo{person}{Lianmin Zheng}, \bibinfo{person}{Wei-Lin Chiang}, \bibinfo{person}{Ying Sheng}, \bibinfo{person}{Siyuan Zhuang}, \bibinfo{person}{Zhanghao Wu}, \bibinfo{person}{Yonghao Zhuang}, \bibinfo{person}{Zi Lin}, \bibinfo{person}{Zhuohan Li}, \bibinfo{person}{Dacheng Li}, \bibinfo{person}{Eric~P. Xing}, \bibinfo{person}{Hao Zhang}, \bibinfo{person}{Joseph~E. Gonzalez}, {and} \bibinfo{person}{Ion Stoica}.} \bibinfo{year}{2023}\natexlab{}.
\newblock \bibinfo{title}{Judging LLM-as-a-Judge with MT-Bench and Chatbot Arena}.
\newblock
\newblock
\showeprint[arxiv]{2306.05685}~[cs.CL]
\urldef\tempurl%
\url{https://arxiv.org/abs/2306.05685}
\showURL{%
\tempurl}


\end{thebibliography}


\end{document}